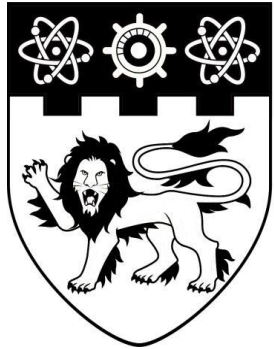

# Speech-preserving active noise control: a deep learning approach in reverberant environments

DAI SHUNING
SCHOOL OF ELECTRICAL AND ELECTRONIC ENGINEERING
2026

# Speech-preserving active noise control: a deep learning approach in reverberant environments

DAI SHUNING

SCHOOL OF ELECTRICAL AND ELECTRONIC ENGINEERING

A DISSERTATION SUBMITTED IN PARTIAL FULFILMENT OF
THE REQUIREMENTS FOR THE DEGREE OF
MASTER OF SCIENCE IN
SIGNAL PROCESSING AND MACHINE LEARNING
2026

# Statement of Originality

I hereby certify that the work embodied in this thesis is the result of original research, is free of plagiarised materials, and has not been submitted for a higher degree to any other University or Institution.

7 April 2026

. . . . . . . . . . . . . . . . . .

Date

*Dai Shuning*

. . . . . . . . . . . . . . . . . . . . . . . . . . .

Dai Shuning

# Supervisor Declaration Statement

I have reviewed the content and presentation style of this thesis and declare it is free of plagiarism and of sufficient grammatical clarity to be examined. To the best of my knowledge, the research and writing are those of the candidate except as acknowledged in the Author Attribution Statement. I confirm that the investigations were conducted in accord with the ethics policies and integrity standards of Nanyang Technological University and that the research data are presented honestly and without prejudice.

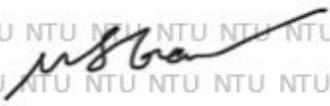

7 April 2026

. . . . . . . . . . . . . . . . . .                            . . . . . . . . . . . . . . . . . . . . . . . . . . . .

        Date                                                Gan Woon Seng

# Authorship Attribution Statement

This thesis **does not** contain any materials from papers published in peer- reviewed journals or from papers accepted at conferences in which I am listed as an author.

7 April 2026

. . . . . . . . . . . . . . . . . .

Date

*Dai Shuning*

. . . . . . . . . . . . . . . . . . . . . . . . . . .

Dai Shuning

# Table of Contents

# Abstract

Traditional Active Noise Control (ANC) systems are mostly based on FxLMS algorithms, but such algorithms rely on linear assumptions and are often limited in handling broadband non-stationary noise or nonlinear acoustic paths. Not only that, the traditional method is used to eliminating all signals together, and noise reduction often accidentally damages the voice signal and affects normal communication. To tackle these issues, this study proposes a speech preserving deep learning ANC system, which aims to achieve stable noise reduction while effectively retaining speech in a complex acoustic environment.

This study builds an end-to-end control architecture, the core of which adopts a Convolutional Recurrent Network (CRN). The structure uses the long short-term memory (LSTM) network to capture the time-related characteristics of acoustic signals. Combined with complex spectrum mapping (CSM) technology, the nonlinear distortion problem is effectively solved.In order to retain useful voice while removing noise, this study also designs a special voice retention loss function. This design guidance model selectively retains the target voice while suppressing environmental noise by identifying the characteristics of the spectrum structure. In addition, in order to verify whether the system is effective in real scenes, we use the Image Source Method (ISM) to build a high-fidelity acoustic simulation environment, which also simulates the real reverberation effect.

Experimental results demonstrate that the proposed Deep ANC system achieves significantly better noise reduction than the traditional FxLMS algorithm, especially for non-stationary noises like crowd babble. Meanwhile, PESQ and STOI based evaluations confirm that the system preserves both the naturalness and intelligibility of the target speech.

# Acronyms

| | |
|---|---|
| ANC | Active Noise Control |
| CRN | Convolutional Recurrent Network |
| CNN | Convolutional Neural Network |
| LSTM | Long Short-Term Memory |
| FxLMS | Filtered-x Least Mean Square |
| STFT | Short-Time Fourier Transform |
| ISTFT | Inverse Short-Time Fourier Transform |
| RIR | Room Impulse Response |
| ISM | Image Source Method |
| SNR | Signal-to-Noise Ratio |
| NR | Noise Reduction |
| PESQ | Perceptual Evaluation of Speech Quality |
| STOI | Short-Time Objective Intelligibility |
| PSD | Power Spectral Density |
| MSE | Mean Squared Error |
| ELU | Exponential Linear Unit |
| RT60 | Reverberation Time 60dB |

# Acknowledgement

First of all, I wish to extend my heartfelt appreciation to my supervisor, Prof. Gan Woon Seng. Thank you for your essential guidance, tolerance, and support throughout the entire research endeavor. Your expertise in signal processing and deep learning was pivotal in the development of this thesis.

I would also like to express my sincere gratitude to Dr. Luo Zhengding and Mr. Yang Ziyi. During the implementation phase of the project, they not only furnished me with valuable technical guidance but also extended altruistic assistance and support regarding the experimental details.

I am also deeply appreciative of the School of Electrical and Electronic Engineering at Nanyang Technological University. The computing resources and stimulating academic environment supplied by the institution were vital for this research.

Finally, I want to thank my family and friends for their encouragement and understanding during my master's studies.

<div style="text-align: right;">

Dai Shuning

7 April 2026

</div>

# Symbols

| | |
|---|---|
| $x(n)$ | Reference signal |
| $y(n)$ | Control signal generated by the ANC controller |
| $d(n)$ | Primary noise signal |
| $e(n)$ | Residual error signal |
| $a(n)$ | Anti-noise signal |
| $s(n)$ | Secondary path impulse response |
| $S(z)$ | Secondary path transfer function |
| $P(z)$ | Transfer function of the primary path |
| $W(z)$ | Transfer function of the adaptive controller |
| $X(m, k)$ | Complex STFT of reference signal |
| $X_r(m, k)$ | Real part of the complex spectrum |
| $X_i(m, k)$ | Imaginary part of the complex spectrum |
| $m$ | Time frame index in STFT |
| $k$ | Frequency bin index in STFT |
| $\theta$ | Learnable parameters of the deep neural network |
| $\mathcal{L}$ | Loss function used for training |
| $RT_{60}$ | Reverberation time (60 dB decay) |

# List of Figures

# List of Tables

# Chapter 1
# Introduction

This chapter introduces the research background in detail, focussing on the severe challenges faced by low-frequency noise control and the limitations of existing linear adaptive algorithms at the physical level. The content arrangement of this entire chapter is as follows: Section 1.1 expounds the research motivation and deeply analyses the shortcomings of the traditional FxLMS algorithm in nonlinear and non-steady noise processing. And Section 1.2 clarifies the specific objectives and scope of this research, and proposes a speech-preserving ANC solution based on deep learning. Finally, Section 1.3 briefly outlines the organisational structure of the whole paper and the logical relationship between each chapter.

## 1.1 Motivations

The acceleration of industrialization and urbanization has made environmental noise pollution a serious global problem. Whether it is vehicles, aircraft and other vehicles, industrial equipment such as engines and blowers, and even social activities themselves are making noise. This not only lowers the quality of life, but also lays hidden dangers for hearing health. For a long time, the suppression of acoustic noise mainly relies on passive noise control technology, such as sound-absorbing materials and sound insulation barriers. As Hansen pointed out, these passive methods are indeed efficient in dealing with broadband high-frequency noise, but in the face of low-frequency noise, they are often cumbersome, expensive and ineffective [1]. This physical limitation has prompted researchers to explore more effective low-frequency noise control solutions.

The suppression of noise in ANC is achieved through the application of the wave superposition principle. It reduces the noise by producing a signal that is opposite in phase to the original noise. Kuo and Morgan expound its basic principles and emphasize that the technology is particularly good at low-frequency noise processing, which makes up for the lack of passive methods. In addition, they also pointed out that the ANC system has potential advantages in terms of volume and weight [2]. In the past few decades, the adaptive filtering algorithm has become an industry benchmark for ANC applications, among which the filter-x least mean square (FxLMS) algorithm is particularly prominent.

Traditional linear adaptive control algorithms face severe challenges in practical application. As Kuo and Morgan pointed out, secondary paths and noise sources in reality often show nonlinear characteristics, or broadband and non-stationary. The FxLMS algorithm based on linear hypothesis is difficult to effectively model these nonlinear distortions. When dealing with broadband noise, its convergence speed also slows down significantly [2]. In addition, existing ANC systems usually adopt a "one-size-fits-all" strategy to eliminate all sound received by the microphone. In scenarios such as factory workshops or cockpits, this rigid method will accidentally damage useful voice signals, hinder normal communication, and even affect the reception of alarm information. Figure 1.1 intuitively compares the differences between traditional linear ANC and our proposed deep learning ANC in processing voice signals.

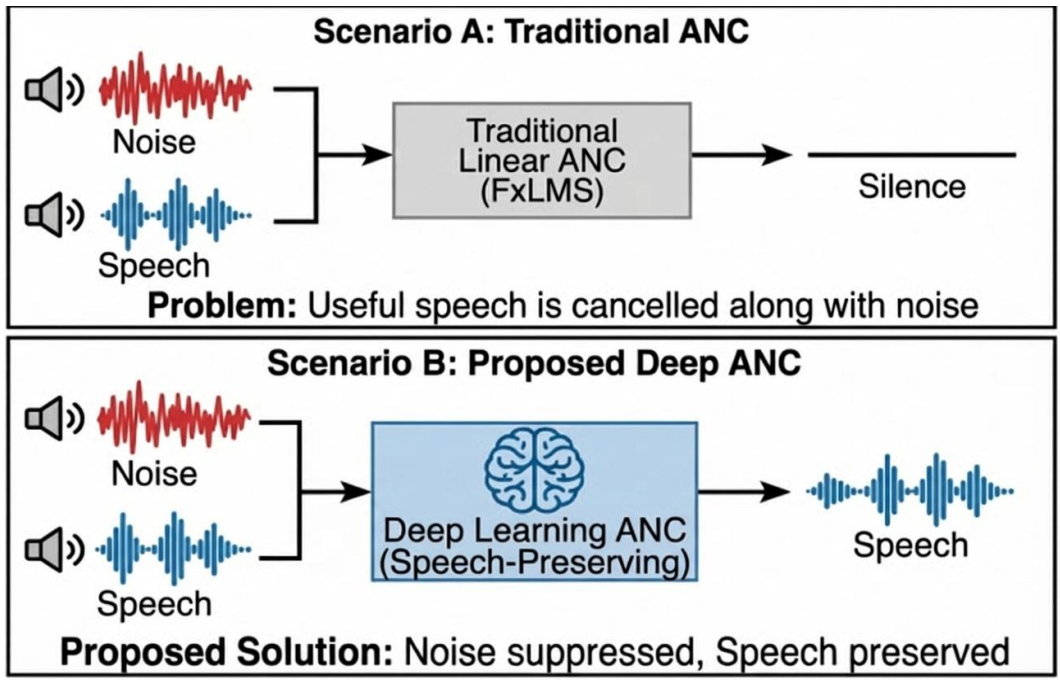

Figure 1.1 Traditional Linear ANC (FxLMS) vs Proposed Deep Learning ANC

*Note: The Figure generated by Google Gemini*

Therefore, both academia and industry urgently need to develop a new generation of intelligent ANC systems. This system should not only be able to adapt to complex acoustic environments and handle nonlinear distortion, but also have selective noise cancellation capability: effectively retain speech while suppressing noise.

## 1.2 Objectives and Scope

In order to break through the limitations of the traditional ANC algorithm in complex sound field and non-stationary noise processing, this study has deeply explored a set of ANC system based on deep learning. The core of the work is to build an end-to-end deep learning framework and examine its robustness in a real reverb environment, as well as the ability to make selectively preserve speech and retain speech. The specific research objectives are as follows:

**Objective 1: Build an end-to-end nonlinear ANC system based on CRN architecture**

In the face of nonlinear acoustic paths, traditional linear adaptive filters (such as FxLMS) are often difficult to do so. Refer to the results of Zhang and Wang [3], this study builds a set of control system based on convolutional recurrent network (CRN). The system uses the causal convolution layer to extract the spectral characteristics of noise and introduces a LSTM unit to capture long-time series of time dependencies. In this way, it establishes a nonlinear mapping from the reference microphone signal to the complex spectrum of the anti-noise signal, so as to effectively deal with the problem of broadband noise control.

**Objective 2: Evaluate the robustness of the system in the real acoustic reverberation environment**

In order to narrow the gap between experimental results and practical applications, this study is no longer limited to simple low-pass filter simulation, but builds a high-fidelity acoustic simulation platform. Using the Image Source Method (ISM) proposed by Scheibler et al. [4], we simulate a physical space with real reverberation time (RT60 = 0.3s) and multi-path propagation effect. On this basis, the performance differences of the model in the ideal noise-free environment and the real reverberation environment are systematically compared, and the robustness of the deep learning model in the face of phase delay and environmental interference is emphasized.

**Objective 3: Benchmark industry standard algorithm**

To provide an objective assessment of the proposed method, this study implements the classic FxLMS algorithm as a baseline for comparison. We have carried out systematic tests on five representative noise sources extracted from the NOISEX-92 database in a unified physical simulation environment. These noise sources cover a variety of types, from low-frequency stationary noise (such as Volvo) to broadband non-stationary noise (such as Factory1, Babble), aiming to comprehensively examine the robustness of the algorithm under diverse acoustic characteristics. By comparing the level of noise reduction (NR) and the steady-state error after convergence, we quantitatively demonstrate the superiority of the deep learning method over traditional signal processing in complex noise scenarios.

**Objective 4: Develop a Deep Learning Strategy for Speech Preservation**

In view of the problem of speech damage caused by full muffled noise reduction in the traditional ANC system, this study proposes a set of improved training strategies and loss function mechanisms. By optimizing the loss function and using pure voice as the constraint target, the training model automatically differentiates and retains the energy of the audio segment, while suppressing background noise. This method aims to resolve the contradiction between noise reduction and voice communication, so as to improve the comprehensibility of voice in noisy environments.

## 1.3 Thesis Organization

This paper contains a total of six chapters, with the content of each described below:

- Chapter 1: Introduction. This chapter first outlines the background and motivation behind the study, then defines the main objectives and scope of the research, and finally briefly describes the organizational structure of the whole paper.

- Chapter 2: Literature Review. This chapter reviews the theoretical basis of traditional active noise control (ANC), particularly the FxLMS algorithm. It also summarizes the current applications of deep learning in audio signal processing and the key techniques used in acoustic environment simulation.

- Chapter 3: Methodology and System Design. This chapter elaborates the core design of the system in detail. These include the architectural details of the convolutional recurrent network (CRN), the high-fidelity acoustic environment modelling method based on *pyroomacoustics*, and the loss function and training strategy specially designed for speech retention tasks.

- Chapter 4: Experimental Setup. This chapter describes the specific implementation details of the experiment, covering the data preprocessing process and the parameter configuration of the baseline model (FxLMS). And define key evaluation indicators, including noise reduction (NR), speech perception quality (PESQ) and short-term objective intelligibility (STOI).

- Chapter 5: Results and Analysis. This chapter shows and analyses the

experimental results in depth. The content includes the comparison between the ideal environment and the real reverb environment, the performance difference between Deep ANC and the traditional FxLMS algorithm, and the qualitative and quantitative evaluation of the selective cancellation and speech retention effect of the system.
- Chapter 6: Conclusion and Future Work. This chapter concludes the key academic contributions and research results of the paper. And put forward suggestions for future research on low-latency hardware deployment and multi-channel system expansion.

# Chapter 2
# Literature Review

This chapter serves to comprehensively review the theoretical basis and cutting-edge progress related to this research. We have systematically sorted out the evolution from traditional signal processing to modern deep learning technology. This chapter is structured as follows. First, Section 2.1 expounds the basic principle of traditional ANC, focussing on the analysis of the classic FxLMS algorithm and its limitations. And Section 2.2 discusses the application of deep learning in audio signal processing, focusing especially the rise of CRN. Subsequently, Section 2.3 presents an in-depth study of the simulation methodology of the real acoustic environment, and introduce the role of Image Source Method (ISM) in the construction of high-fidelity sound field. Finally, Section 2.4 introduces the concept of selective noise control, and the unique challenges faced by this study in physical sound field control are clarified by comparing speech enhancement (SE) tasks.

## 2.1 Traditional Active Noise Control

The ANC system uses the sound wave superposition principle to reduce noise. It generates a sound wave with an amplitude equal to the original noise but in opposite phase. The two kinds of sound waves interact in the sound space to produce phase cancellation interference, thus canceling the noise.

In the past few decades, the ANC system has mainly relied on adaptive filtering theory. Among various algorithms, the FxLMS algorithm has become an industry benchmark with the advantages of simple structure and high stability. As mentioned by Kuo and Morgan, the FxLMS algorithm introduces a compensation mechanism: Before updating the control weight, filter the reference signal to compensate for the effect of the electroacoustic path between the secondary speaker and the error microphone (i.e. the secondary path) [2]. This mechanism ensures that the system can remain stable even if there is a physical delay. In addition, Elliott builds a detailed mathematical framework for active control, and deeply analyzes the performance limits of the feedforward and feedback architecture [7].

To achieve a better balance between the steady-state error and convergence speed, researchers have proposed several improved adaptive algorithms based on the classic FxLMS. For example, the normalized filtered-x least mean square (FxNLMS) algorithm employs a variable step length mechanism, which significantly improves the system's robustness against power fluctuations of the input signal [8]. In addition, the ANC system using recursive least squares (RLS) has also been developed, which greatly accelerates the initial convergence speed [8]. At the same time, considering that secondary paths in practical applications are often time-varying (such as temperature changes or microphone displacement), online secondary path modeling (OSPM) technology has been widely studied. Akhtar et al. proposed an improved OSPM structure to ensure that the system can track secondary path changes in real time and always maintain the stability of the controller during the process of noise reduction [9].

Driven by these improved algorithms, the traditional ANC system has achieved great commercial success. Their application scenarios have gradually expanded from the early single-channel pipeline noise cancellation to complex multi-channel 3D space noise control. ANC technology has also become a standard function of high-end consumer electronic products (such as ANC headphones). This technology is also deeply integrated into the automotive industry, actively suppressing engine and road noise in the car cabin, so as to improve the acoustic comfort of passengers [10].

However, traditional linear algorithms still face inherent limitations in practical application. First, the FxLMS algorithm relies on the linear hypothesis, meaning that the acoustic path and the noise source are linear systems. However, in reality, hardware components such as speakers and amplifiers, especially driven by high volume, often introduce nonlinear distortion. Second, as Kuo and Morgan pointed out, the convergence speed of linear adaptive filters tends to be slow when dealing with broadband or colored noise [2]. Therefore, in the face of dynamic changes in the non-stationary noise environment, these traditional approaches are often ineffective at tracking.

## 2.2 Deep Learning for Audio Signal Processing

In order to break through the bottleneck of linear signal processing, researchers turned their attention to deep learning. Purwins et al. sorted out the evolution of this technology, from a simple deep neural network (DNN) to a more complex convolutional neural network (CNN) and a recurrent neural network (RNN) [11].

Specifically, CNN is good at extracting spectral features, while RNN performs well in the modeling of long-term dependencies.

The strong timing dependence of audio signals has attracted continuous attention to LSTM and the broader RNN architecture in the ANC field. Pike and Cheer systematically compared the control performance of three neural network architectures in feedforward nonlinear ANC systems: MLP, Elman RNN and LSTM [12]. Experimental results show that when dealing with nonlinear distortion in the primary path, LSTM shows the optimal control effect due to its stronger long-term memory ability, and its calculation cost is acceptable in practical application. Kwon et al. further presented an ANC controller based on LSTM. The simulation results demonstrate that, compared to the conventional FxLMS algorithm, the LSTM controller exhibits superior noise reduction across narrowband, broadband, and pulse noise environments [13]. In addition to simulation research, LSTM has also been successfully applied to practical engineering scenarios. For example, Cha et al. have developed CsNNet, a deep learning ANC controller that integrates LSTM and attention module, which achieves the best noise reduction effect on 17 different construction mechanical noises [14]. Huang et al. used LSTM to construct a predictive model specifically for side window wind noise in vehicles, and realized high-precision noise prediction in the 100-500 Hz band [15].

On the basis of the above research, hybrid architecture has gradually become the mainstream of research in the field of ANC. Cao et al. proposed the enhanced LSTM-ANC (ELSTM-ANC-OSPM) framework, which combines LSTM with online secondary path modeling to achieve more stable noise reduction performance than traditional methods in a time-variable environment [16]. In the multi-channel ANC scenario, Zhu et al. designed a hybrid network based on CNN-LSTM and proposed a new adaptive sound zone control strategy [17]. Then they further proposed a hybrid system combining adaptive self-loading FxLMS integrated with a CNN-GRU network for the time-variable noise environment. Among them, GRU, as a lightweight variant of LSTM, reduces the computing load while maintaining the timing modeling ability. Experiments show that the robustness and tracking performance of this method under time-variable noise are better than that of traditional Deep ANC network [18]. In addition, Jabez et al. introduced CNN-BiLSTM network for real-time complex noise reduction to capture more comprehensive timing dependencies through bidirectional LSTM (BiLSTM). However, it should be pointed out that BiLSTM violates causal constraints due to the use of future information, and its application is mainly limited to offline analysis [19].

A milestone breakthrough in this field is the introduction of convolutional recurrent network (CRN). Tan and Wang designed the CRN architecture for real-time voice enhancement tasks [20]. The model employs a Convolutional Encoder-Decoder (CED) to capture high-level features and embeds long-term and short-term memory (LSTM) units to process timing dynamics. It is worth noting that the architecture is causal, that is, it does not rely on future information and is very suitable for real-time applications. On this basis, Zhang and Wang put forward the concept of Deep ANC, which proves the feasibility of using supervised learning to directly establish reference signal and anti-noise signal mapping [21]. Subsequently, they expanded it into a complete set of deep learning ANC frameworks in their journal papers by Zhang and Wang [3]. This method does not need to explicitly identify secondary paths, and is superior to FxLMS in dealing with nonlinear distortion.

To obtain rapid and stable noise reduction without the slow convergence issue of adaptive algorithms, Shi et al. introduced a CNN-based Selective Fixed-Filter Active Noise Control (SFANC) method that chooses suitable pre-trained control filters according to the characteristics of incoming noises, enabling delayless real-time implementation with high robustness [42, 43]. However, the performance of SFANC is fundamentally constrained by the limited number of candidate filters. In order to break through this limitation, Luo et al. proposed the generated fixed filter active noise control (GFANC) technology. The technology decomposes a sub-control filter from a single broadband filter through a lightweight CNN adaptive combination to generate a control filter that matches the noise [44, 45]. On the basis of GFANC, the researchers have developed a number of extension schemes to improve robustness and practicality, including GFANC variants that use the timing correlation between adjacent noise frames, and the unsupervised learning and reinforcement learning GFANC method without marking noise data [46, 47, 48, 49].

In recent years, relevant research has continued to deepen. Zhang and Wang proposed the Attention Recurrent Network (ARN) to make the system respond faster and have lower latency [22]. Park et al. explored a hybrid system (HAD-ANC), which combines the stability of traditional adaptive filters with the flexible modeling power of deep neural networks to specifically deal with complex environments [23]. In addition, Liu et al. systematically summarized the ANC algorithm based on neural network in the review, sorted out the research progress from the two dimensions of network architecture and training method, and provided a comprehensive reference for the application of neural networks such as LSTM in ANC [24]. A recent review systematically links traditional and learning-based active noise control methods [25]. And Chen et al. emphasized emerging trends in the review, such as the application of Transformer architecture to noise control, which pointed out the direction for the future development of intelligent ANC systems [26]. Lately, Yaish et al. and Mishaly et al. further expanded this direction and explored the application of Transformer-Mamba hybrid architecture and optimization-driven loss functions in active speech enhancement and active speech elimination respectively [27, 28].

## 2.3 Acoustic Environment Simulation

One of the difficulties of ANC research is how to shorten the gap between theoretical simulation and physical reality. Early research often assumed ideal free field conditions, or simulated acoustic paths with a simple low-pass filter. However, the real acoustic environment is far more complex than this, full of significant reverberation and multi-path propagation effects.

In order to simulate the real room acoustic effect, the Image Source Method (ISM) proposed by Allen and Berkley is still the classic and most widely used method [29].

This method simulates wall reflection by building a virtual mirror sound source to calculate the room pulse response (RIR). In modern research, Scheibler et al. developed pyroomacoustics, a Python toolkit that efficiently implements ISM [4]. With the help of this tool, researchers can customise the room size, wall sound absorption coefficient and microphone array layout to generate high-fidity RIR. This study uses this tool to evaluate the robustness of the Deep ANC model in the reverberation environment ($RT_{60} = 0.3s$). Compared with the ideal simulation, the verification process is more rigourous.

It should be pointed out that the reverberation environment will reduce the maximum noise reduction capacity of the ANC system and make a significant difference in the convergence speed of noise at different frequencies [30]. In recent years, researchers have further explored the impact of reverb on deep learning ANC systems. For example, Wang et al. proposed a CNN-based directional SFANC method, which achieves a better noise reduction effect compared with traditional adaptive algorithms in the reverberation environment, and verifies the robustness of the deep learning model in a complex acoustic environment [31]. Therefore, evaluating the performance of the ANC algorithm under reverberation conditions is an important link to verify its practical application potential.

## 2.4 Speech Preservation and Selective ANC

Traditional ANC systems usually adopt a comprehensive elimination strategy, regardless of noise or useful information. However, in practical application, people prefer to selectively eliminate noise while retaining key signals such as voice or alarm.

Xiao et al. explored the spatially selective ANC system, which can create a silent area in a specific location and let the sound pass through other areas [32]. Although spatial selection is very useful, the demand for spectral-based selective control is growing, that is, to distinguish signals according to their spectral characteristics (e.g., speech vs. noise) rather than location. In the Deep ANC framework proposed by Zhang and Wang, the network learns to act as a semantic filter through supervised learning. The training goal is no longer to output mute, but to output pure voice [3,21].

It should be pointed out that the realization of this selective cancellation requires accurate phase control, because ANC relies on the phase cancellation interference of sound waves. The early deep learning audio processing model mainly focused on amplitude spectrum estimation, ignoring the importance of phase. Recent work indicates that phase information plays a key role in producing high-fidelity anti-noise signals and maintaining perceived speech quality. Therefore, complex spectrum mapping (CSM) has emerged as a better method. Through direct estimation of both the real and imaginary spectral components, CSM effectively avoids the thorny phase estimation problem and enhances the amplitude and phase information at the same time [33,35]. The effectiveness of CSM has been verified in challenging acoustic tasks. For example, the DCCRN model using complex CRN architecture and CSM won the first place in the real-time category of the Interspeech 2020 Deep

Noise Suppression Challenge [34]. This further proves that the integration of CSM into the CRN-based ANC framework is a natural and reasonable evolution to meet the strict phase requirements of acoustic cancellation.

This speech-preserving ANC method must be distinguished from the traditional speech enhancement (SE) task [16, 20]. Although both are committed to improving voice comprehension in noisy environments, ANC faces unique physical challenges that the SE model cannot solve:

First, real-time constraints. ANC must generate an anti-noise signal within the delay of sound wave propagation (usually less than 10 milliseconds) to achieve effective cancellation. In contrast, SE algorithms can usually tolerate longer processing delays and even allow offline processing.

Second, physical superposition. ANC realises noise reduction through the phase cancellation interference of sound waves in the air. SE is purely to filter and denoise audio recordings in the digital domain.

Third, secondary path compensation. The Deep ANC controller must consider the electroacoustic transfer function of the secondary path. The path extends from the secondary speaker to the error microphone. The network must learn to generate a specific control signal, which can only produce a cancellation effect after propagated through this physical path. The SE model does not need to model this physical propagation.

This research aims to overcome the problem of physical acoustic path that cannot be solved by the pure SE model. It focusses on how the CRN architecture can selectively attenuate background noise in a real and reverberant acoustic environment while maintaining the comprehensibility of speech.

# Chapter 3
# Methodology and System Design

This chapter describes the design and implementation of a deep learning based active noise control system. In the face of nonlinear distortion and broadband non-stationary noise, traditional linear algorithms are often unable to do so. To this end, this study has built a set of end-to-end deep learning frameworks. The content of the whole chapter is arranged as follows: Section 3.1 uses the concept of digital signal processing to mathematically modelle the ANC problem. And Section 3.2 details the convolutional recurrent network (CRN) architecture adopted by this system.  Section 3.3 Explains the real acoustic environment simulation scheme based on the Image Source Method (ISM). Section 3.4 focusses on the specific training strategies and loss functions designed for the realisation of voice retention.

## 3.1 Problem Formulation

The basic principle of ANC lies in the use of phase cancellation interference of sound waves. It produces an anti-noise signal with an amplitude equal to the original noise but in opposite phase, thus cancelling the original noise [1]. In order to effectively apply deep learning to ANC tasks, the physical acoustic process must be strictly defined within the framework of digital signal processing.

## 3.1.1 Signal Model of ANC Systems

This research is mainly aimed at the single-channel feedforward ANC system. This kind of system usually consists of three key parts: reference path, control path and secondary path. The system architecture is illustrated in Figure 3.1. The main hardware includes a reference microphone, a secondary speaker for cancelling sound waves, and an error microphone for monitoring errors.

We use $n$ to represent the discrete time point, and $x(n)$ represents the reference signal captured by the reference microphone. As shown in Figure 3.1, the controller processes this signal to generate the control signal $y(n)$, which then drives the secondary loudspeaker.

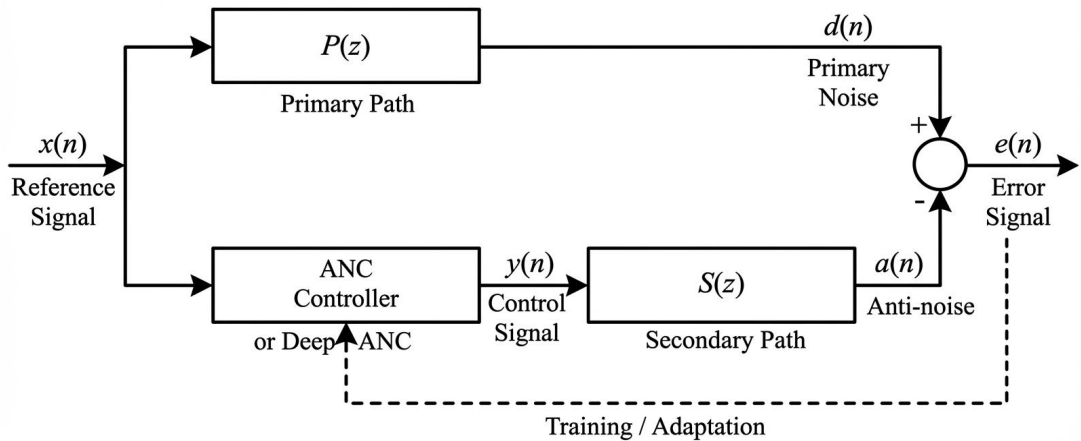

Figure 3.1 Single-channel Feedforward Deep ANC System

According to the sound wave superposition principle, the residual error $e(n)$ captured by the microphone represents the linear sum of the primary noise and the secondary anti-noise. According to the definition of Zhang and Wang [3], under the ideal assumption of ignoring the nonlinearity of the speaker, $e(n)$ can be expressed as:

$$e(n) = d(n) + a(n) \tag{3.1}$$

Among them, $d(n)$ represents the noise component of the original noise through the propagation path (i.e. the primary path $P(z)$) to the error microphone, while $a(n)$ is the anti-noise signal generated by the secondary speaker.

It must be pointed out that the control signal $y(n)$ does not directly become the anti-noise signal $a(n)$. It needs to go through a physical acoustic path and a series of electronic components first. This comprehensive path is called a secondary path and is written as $S(z)$. It covers digital-to-analogue converter (D/A), reconstruction filter, power amplifier, and the acoustic channel between the speaker and microphone, followed by the A/D converter [2]. Therefore, the actual anti-noise signal $a(n)$ is actually the linear convolution of the control signal $y(n)$ and the secondary path pulse response $s(n)$:

$$a(n) = s(n) * y(n) \tag{3.2}$$

The formula (3.2) is substituted into (3.1), and the error equation of the ANC system can be rewritten as:

$$e(n) = d(n) - s(n) * y(n) \tag{3.3}$$

The core task of the ANC system is to find the best control strategy to generate $y(n)$. For pure noise reduction tasks, the goal is to minimise the energy of the residual error signal $e(n)$. In a voice retention task, the goal is to make the error signal as close as possible to the specific target voice signal.

Traditional ANC methods, such as the classic FxLMS algorithm, usually model the controller as a linear adaptive filter $W(z)$. Its control signal is generated by linear convolution: $y(n) = w(n) * x(n)$ [2]. However, in reality, hardware (such as speakers) often has nonlinear distortion, and the broadband noise environment is complex and variable, and linear filters are often unable to do so.

In order to break through this limitation, this study adopts Deep Neural Network (DNN) as the controller. We have reconstructed the ANC problem into a supervised learning task. The goal is to train a nonlinear mapping function $f_\theta(\cdot)$, where $\theta$ represents the network parameter. Under this framework, the generation of control signals is no longer limited by linear convolution, but refers to the nonlinear function of the signal sequence:

$$y(n) = f_\theta(x(n), x(n-1), ..., x(n-T)) \tag{3.4}$$

Through this method, the deep neural network can learn the complex nonlinear relationship between the reference signal $x(n)$ and the ideal control signal. This enables the system to show better noise reduction performance than traditional linear algorithms in the face of complex acoustic environments with path delay and nonlinear effects [3].

### 3.1.2 Spatial and Spectral Selectivity

Traditional ANC systems usually adopt an undifferentiated suppression strategy, and its core goal is to minimise the mean square of the error signal ($E[e^2(n)] \to 0$). However, in actual scenarios such as wearing headphones for face-to-face communication, the situation is often more complicated. The sound received by the error microphone not only contains the background noise that needs to be eliminated, but also the target voice that the user wants to hear.

In response to this problem, this study introduces the concept of spectral selectivity. This inspiration comes from the spatial spectral proposed by Xiao et al. in the multi-channel ANC system [7]. They use the beamforming technology of the microphone array to retain the sound in a specific direction through spatial filtering. Unlike this, this study uses the powerful feature extraction ability of deep neural network (DNN), which is no longer limited to spatial position, but distinguishes and processes mixed signals according to spectral structures and signal content. From a mathematical point of view, we decompose the original interference signal $d(n)$ at the error microphone into two independent parts:

$$d(n) = v(n) + s_{\text{target}}(n) \qquad (3.5)$$

Among them, $v(n)$ represents the environmental noise (such as engine sound or fan noise) that needs to be eliminated, while $s_{target}(n)$ represents the target voice signal that needs to be retained. Figure 3.2 intuitively shows the difference between the traditional full elimination strategy and the semantic selective strategy proposed in this study in ANC.

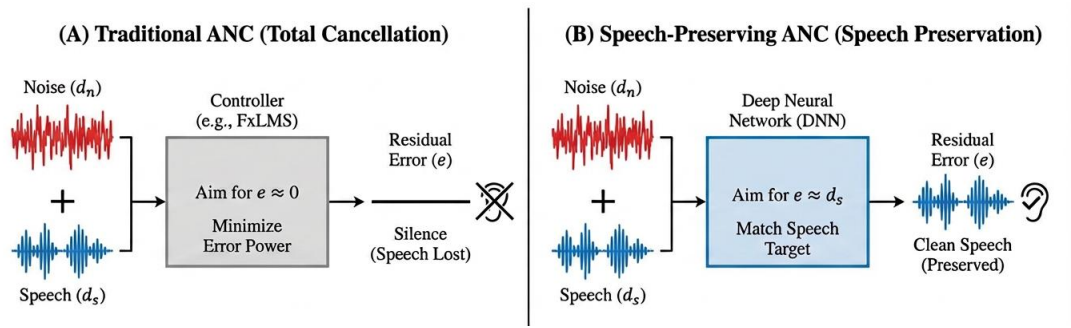

Figure 3.2 Comparison of traditional total cancellation and the proposed semantic

*Note: The Figure generated by Google Gemini*

In this scenario, the goal of the ideal ANC system is no longer to make $e(n) = 0$, but to act as a selective filter. Our goal is to make the residual error signal as close as possible to the target voice signal, that is:

$$e(n) \approx s_{\text{target}}(n) \tag{3.6}$$

This means that the anti-noise signal $a(n)$ generated by the secondary speaker should only offset the noise component $v(n)$ (i.e. $a(n) \approx -v(n)$), and let the target voice $s_{target}(n)$ pass without loss. In order to achieve this acoustic transparency, accurate phase control is essential, and any phase dislocation may inadvertently cancel the voice component. This strict physical requirement is the theoretical basis of the complex spectral mapping strategy (section 3.1.3) and the speech retention loss function (section 3.4) proposed in this paper.

### 3.1.3 Complex Spectral Mapping

Voice and broadband ambient noise are highly non-stationary. If it is only processed in the time domain, it is difficult to effectively extract its spectrum characteristics. Therefore, this study mainly carries out feature extraction and signal reconstruction in the time frequency (T-F) domain.

The traditional speech enhancement method usually adopts time-frequency masking technology, focussing on estimating amplitude spectrum. In the waveform reconstruction stage, these methods often directly reuse the phase information containing noise signals. However, for the ANC task, the accuracy of the phase is crucial. ANC depends on the interference principle of wave. The anti-noise signal must maintain an accurate inverse phase (180° difference) from the original noise to realise phase cancellation interference. If the phase estimation is biased, the system is not only unable to eliminate the noise, but may even cause noise amplification due to phase-length interference [3].

In order to solve this problem, this study adopts the complex spectrum mapping (CSM) strategy proposed by Tan and Wang [35]. Unlike the method of predicting only amplitude spectrum, CSM aims to simultaneously estimate both the real and imaginary components of the target signal in the complex field. This method implicitly models the amplitude and phase information together, thus ensuring the phase consistency in the process of waveform reconstruction.

The procedure can be described as follows. First, the system uses short-time Fourier transform (STFT) to convert the discrete time domain reference signal $x(n)$ to complex spectrum $X(m, k)$:

$$X(m, k) = \sum_{n=0}^{N-1} x(n + mH) w(n) e^{-j\frac{2\pi k n}{N}} \qquad (3.7)$$

Among them, $m$ is the time frame index and $k$ is the frequency index. $N$ stands for FFT points, which is set as $N = 320$ in this study. At a sampling rate of 16 kHz, the frame length corresponds to 20 ms. $H$ represents the length of the jump, set to $H = 160$ (i.e. 10 ms), which means that there is 50% overlap between adjacent frames. $w(n)$ is a Hanning window function.

Then, we decompose the complex spectrum $X(m, k)$ into the real part $X_r(m, k)$ and the imaginary part $X_i(m, k)$. These two parts are entered into CRN as two independent feature channels. The specific processing process is shown in Figure 3.3. The output layer of the network is also designed to output two channels, corresponding to the real part $Y_r(m, k)$ and the virtual part $Y_i(m, k)$ of the predictive control signal respectively.

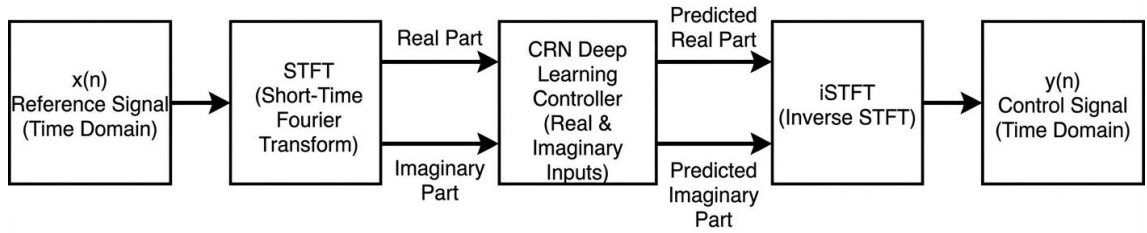

Figure 3.3 Signal processing flowchart based on complex spectral mapping

Finally, use the inverse short-time Fourier transform (iSTFT) to map the predicted complex spectrum back to the time domain and generate the control signal $y(n)$:

$$y(n) = \text{iSTFT}(\hat{Y}_r(m, k) + j\hat{Y}_i(m, k)) \qquad (3.8)$$

Through this end-to-end mapping of complex fields, the Deep ANC system can learn finer acoustic characteristics than traditional amplitude spectrum mapping. This ensures that the generated anti-noise signal can meet the strict requirements of amplitude and phase of noise cancellation [3, 35].

## 3.2 Convolutional Recurrent Network (CRN) Architecture

The ANC system has extremely high requirements for the real-time and non-linear modelling ability of the controller. First of all, limited by the speed of sound wave propagation, the controller must complete the signal processing within a very short millisecond delay. This imposes strict constraints on the causality of the algorithm. Secondly, ambient noise (such as fan sound, crowd sound) usually has complex time-variable spectrum characteristics. This requires that the model must have strong feature extraction ability and timing memory ability. In view of this, this study selects a deep learning architecture optimised for real-time audio processing - convolutional recurrent network (CRN) as the core controller.

The CRN architecture was originally proposed by Tan and Wang to solve the problem of voice enhancement [20]. Subsequently, Zhang and Wang verified their excellent performance in nonlinear ANC tasks [3]. As shown in Figure 3.4, the architecture adopts the classic encoder-decoder structure and cleverly embeds the long-term and short-term memory (LSTM) unit at the bottleneck layer. This design organically combines CNN's advantages in local feature extraction and RNN's strengths in long-time series modeling.

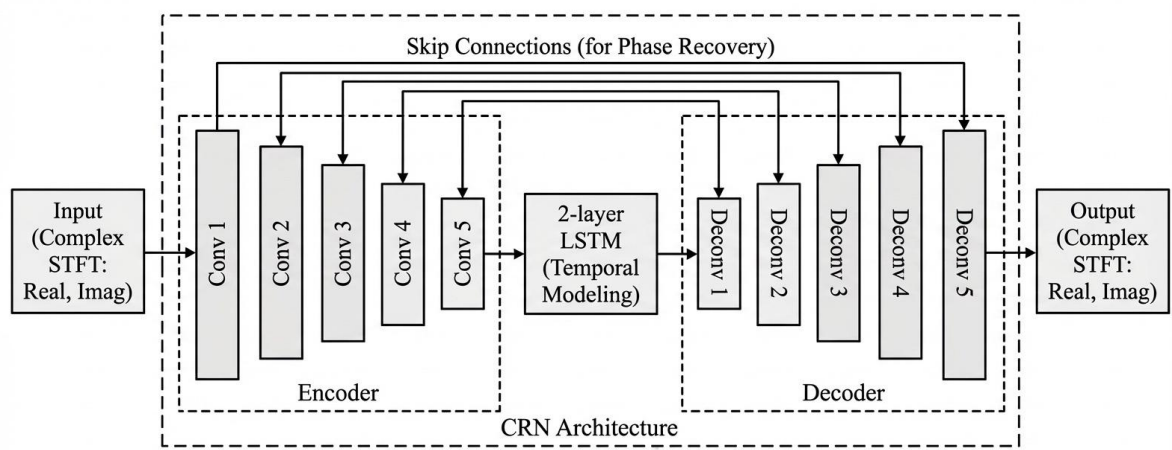

Figure 3.4 CRN Overall Architecture Diagram

## 3.2.1 Convolutional Encoder and Causal Feature Extraction

The core task of the encoder is to compress the complex spectrum of high-dimensional input into a compact advanced acoustic feature representation. The encoder designed in this study contains five stacked two-dimensional convolutional layers (Conv2d).

- Spectrum dimension reduction and compression:

In the direction of the frequency axis, the step length of the convolution kernel is set to 2. This means that every time the data passes through a layer, the frequency dimension of the feature map will be halved, and the number of channels will be doubled. This design not only effectively eliminates the redundant information in the spectrum, but also forces the network to learn the local spectrum structure with translation invariance. This feature is particularly critical for capturing features such as engine noise harmonic textures [36].

- The realisation of causal convolution:

Meeting the causal constraints of the ANC system is the top priority of design. The traditional image convolution kernel uses the information before and after the current moment at the same time. In order to ensure that the system only uses current and past information, we have adopted an asymmetric zero filling strategy. Specifically, only fill in the past direction of the timeline, not in the future direction. This causal convolution design eliminates non-causal delay from the physical level and lays a theoretical foundation for real-time deployment [20].

- Selection of activation function:

With the exception of the output layer, all convolutional layers employ exponential linear units (ELU) as the activation function. Clevert et al. pointed out that compared with ReLU, ELU shows stronger robustness when processing negative input [37]. It can not only accelerate the convergence of the deep network, but also effectively alleviate the problem of gradient disappearance. This is particularly important considering that audio spectrum data often contains negative values.

### 3.2.2 Temporal Modeling Bottleneck

After being compressed by the encoder, the feature sequence enters the bottleneck layer of the network. Environmental noise (such as fan rotation or crowd noise) usually shows a significant long-term correlation. Pure convolutional networks are often unable to catch this kind of time evolution pattern.

- Linear Projection Layer:

The acoustic characteristics of the encoder output are up to 1024 dimensions, and the direct input of LSTM will generate a huge amount of parameters, which brings serious challenges to hardware video memory and real-time. Considering the strict requirements of the ANC system for real-time processing of low latency, and seeking the best balance between limited computing resources and model performance, this study introduces a linear projection layer in the code implementation. This layer maps 1024-dimensional feature compression to 256-dimensional as the input of LSTM.

- LSTM Units:

We have deployed two layers of stacked LSTM on the bottleneck layer. With the help of its gate control mechanism, LSTM can effectively remember the key historical noise modes in long time series. This ability enables it to accurately predict the future state of non-stationary noise.

### 3.2.3 Decoder and Signal Reconstruction

The decoder's task is to gradually restore the abstract temporal features output by the LSTM back to a complex spectrum with the same size as the input. This is done to generate the control signal.

- Transposition convolution:

The decoder consists of five transposed convolutional layers, and its structure is strictly symmetrical with the encoder. Through the upsampling operation, the frequency dimension of the feature map can be restored layer by layer.

- Skip Connections:

To address the issue of information loss in deep networks, this study draws on the design concept of U-Net [38]. We introduced skip connections between matching layers of the encoder and decoder. This mechanism allows the high-resolution details extracted by the encoder to be directly spliced into the input of the decoder. The skip connections effectively retains the tiny phase details in the original signal, which significantly improves the waveform fidelity of the anti-noise signal.

- Linear Output:

The last layer of the decoder does not use the nonlinear activation function, but directly outputs the predicted real and imaginary spectrum through linear regression. Subsequently, use ISTFT to synthesise the time domain control signal $y(n)$.

Table 3-1 summarises the detailed configuration of the proposed CRN architecture.

| Module | Layer Details | Kernel (T, F) | Stride (T, F) | Channels | Activation |
|---|---|---|---|---|---|
| Encoder | 5 × Conv2d | 2 × 3 | (1,2) | 16 → 32 → … → 256 | ELU |
| Bottleneck | Linear Projection | - | - | 1024 → 256 | Linear |
|  | 2 × LSTM Layers | - | - | Hidden Size: 256 | Tanh / Sigmoid |
| Decoder | 5 × Transposed Conv2d | 2 × 3 | (1,2) | 128 → 64 → 32 → 16 → 2 | ELU / Linear* |

Table 3-1 Detailed Layer Configuration and Hyperparameters of CRN Architecture

*Note:The final output layer (2 channels) adopts linear activation to generate complex spectrum (real and virtual). The Skip Connections are applied between the corresponding layer of the encoder and the decoder.*

## 3.3 Acoustic Environment Simulation

In order to strictly evaluate the robustness of the Deep ANC system in scenarios close to reality, this study goes beyond the idealised model. Based on the principle of physical acoustics, we simulate the real acoustic environment by generating room pulse response (RIR). The next section will elaborate on the simulation algorithm, the geometric configuration of the virtual environment and the data synthesis process.

### 3.3.1 The Image Source Method (ISM)

In order to efficiently and accurately generate the RIR of rectangular rooms, this study adopts the Image Source Method (ISM). Originally proposed by Allen and Berkley [29], this method has become a standard means of room acoustic modelling. The core idea of ISM comes from geometric acoustics. It simplifies the sound reflection problem in the closed room to the problem of the propagation of multiple sound sources in the open space (free field). For the sound source placed in a rectangular room, each wall reflection can be simulated by a mirror sound source located outside the room. These virtual sound sources act as mirror images of the original source with respect to the wall.

In this study, We utilized the *pyroomacoustics* library to implement the ISM simulation. This is a comprehensive Python library developed by Scheibler et al. specifically for room audio simulation [4]. With the help of this tool, we can accurately control the physical size and sound-absorbing parameters of the room. This makes the generated RIR have a very high sense of realism and can accurately capture early reflection and late reverberation.

## 3.3.2 Simulation Setup and Configuration

To build a representative and generalizable acoustic environment, we simulated a rectangular room with dimensions of $4m \times 3m \times 2.5m$. This size roughly corresponds to the interior space of a small office or a large vehicle. Regarding acoustic characteristics, the reverberation time ($RT_{60}$) is set to 0.3 seconds to simulate a typical indoor environment with moderate acoustic treatment, such as a carpeted room.. This setting helps to evaluate the robustness of the ANC system in the face of real reverberation interference.

The geometric layout of the ANC system is configured as a typical single-channel feedforward control structure. The specific spatial coordinates of each transducer are listed in Table 3-2, and the overlooking layout is shown in Figure 3.5. The simulation sampling rate is set at 16 kHz, which matches the sampling rate of LibriSpeech and NOISEX-92 data sets, covering the main frequency range of human auditory and voice communication.

| Component | Coordinate $(x, y, z)$ [m] | Description |
|---|---|---|
| Primary Noise Source | $(1.00, 1.50, 1.20)$ | Location of noise source (Co-located with Ref Mic) |
| Reference Microphone | $(1.00, 1.50, 1.20)$ | Capture the reference signal $x(n)$ |
| Error Microphone | $(3.00, 1.50, 1.20)$ | Target position of noise reduction (simulated |

|                  |                  | human ear)            |
|------------------|------------------|-----------------------|
| Secondary Source | (3.05,1.50,1.20) | Speaker playing noise-reducing waves |

Table 3-2 Geometric Configuration of the Simulated ANC System

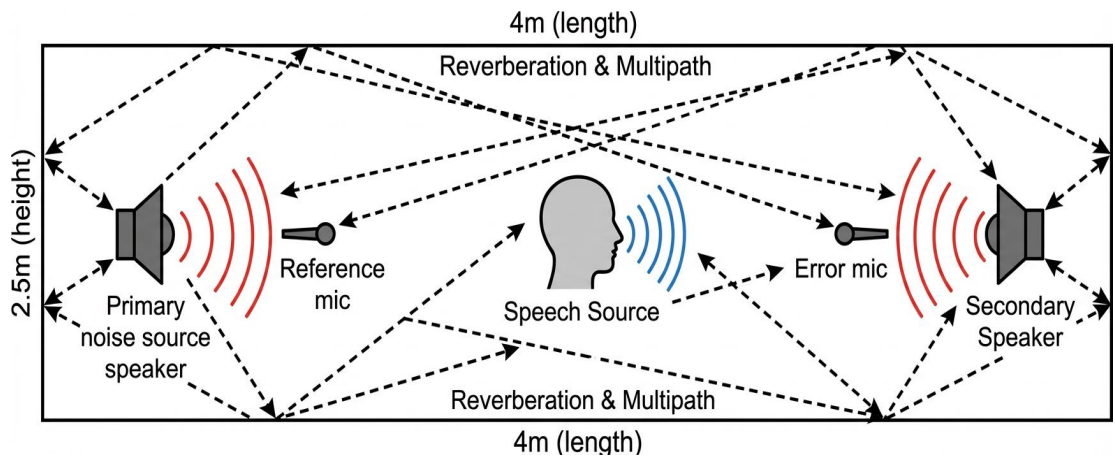

Figure 3.5 Layout of the simulated ANC system

*Note: Figure generated by Google Gemini*

In this configuration, the secondary sound source distance error microphone is only 5 centimetres (3.05m vs 3.00m). This design is designed to simulate typical near-field applications, such as active noise reduction headrests. By minimising the acoustic delay of the secondary path, the layout ensures that the system meets the causal constraints required for broadband noise control. In addition, it also allows the system to generate cancellation signals with high electroacoustic efficiency.

Based on the above geometric parameters and acoustic properties, we use the *pyroomacoustics* library to generate two key room impulse responses (RIR), and truncate its length to 512 points to adapt to the filter design:

- Primary path RIR ($h_{pri}$): Describe the acoustic transfer function of noise from

primary sound source to error microphone.

- Secondary path RIR ($h_{\text{sec}}$): Describe the acoustic transfer function from the secondary speaker to the error microphone.

These two RIRs contain direct sound and reflection sequences with a length of 512 sampling points, which accurately captures the frequency selective attenuation and phase delay characteristics in the simulation room.

### 3.3.3 Data Synthesis Process

In order to train the speech retention ability of the Deep ANC model, we have built a set of data synthesis process based on physical acoustics. This process aims to convert the original reverb-free audio signal into a simulation signal with real room reverberation characteristics.

**1. Original material pretreatment**

The experimental data comes from two standard public data sets:

- NOISEX-92 [5]: Used to provide diversified environmental noise samples ($n_{src}$).

- LibriSpeech [6]: Used to provide pure speech samples ($s_{src}$).

To keep the data consistent, all audio files have been re-sampled to 16 kHz and uniformly normalised.

**2. Reference signal generation**

Refer to the microphone signal $x(t)$ as the input of the network, which needs to simulate the mixed sound field in the real environment. In the simulation setting, we assume that the reference microphone is located in the near-field of the sound source, so the reverberation impact caused by path propagation is temporarily ignored at this step. We adopt a dynamic mixing strategy to linearly superimpose the noise signal with the voice signal:

$$x(t) = n_{src}(t) + \alpha \cdot s_{src}(t) \quad (3.9)$$

Among them, $\alpha$ is the gain coefficient, which is determined by the target signal-to-noise ratio (SNR). SNR random sampling in the range of 0 dB to 10 dB helps to improve the model's adaptability to different noise levels.

**3. Acoustic path simulation**

Using the room impulse responses (RIRs) generated by section 3.3.2, we simulate the physical transmission process of the signal through convolution operation.

- Primary path components: The sound of the arrival error microphone consists of two independent components, which are convolutional with the main path RIR ($h_{pri}$):

1. Primary noise ($d_{noise}$): Noise interference components to be eliminated.

$$d_{noise}(t) = n_{src}(t) * h_{pri}(t) \quad (3.10)$$

2. Primary speech ($d_{speech}$): The target voice component to be retained.

$$d_{speech}(t) = s_{src}(t) * h_{pri}(t) \quad (3.11)$$

- Secondary path ($a(t)$): In the training circuit, the anti-noise signal $y(t)$ generated by the controller must be convolutioned with the secondary path RIR ($h_{sec}$) to simulate the electroacoustic propagation process from the secondary speaker to the error microphone. This produces an anti-noise signal $a(t)$ that actually acts on the sound field:

$$a(t) = y(t) * h_{\sec}(t) \tag{3.12}$$

4. **Target signal definition**

In the speech retention task, the control goal is to eliminate the noise component while completely retaining the natural voice transmitted through the room.. Therefore, we define the trained target signal $s_{target}(t)$ as the voice component after transmission through the main path:

$$s_{target}(t) = d_{speech}(t) = s_{src}(t) * h_{pri}(t) \tag{3.13}$$

By setting the target as the voice $d_{speech}(t)$ after physical transmission, the loss function drives the network to generate a specific anti-noise signal $a(t)$, so that it only interferes with the noise component $d_{noise}(t)$ (i.e. $a(t) \approx - d_{noise}(t)$), and remains transparent to the speech component. This mechanism enables the CRN model to learn the acoustic characteristics of physical space and the inverse characteristics of secondary paths.

## 3.4 Speech Preservation Strategy and Loss Function

In order to suppress background noise while retaining the target voice signal, this study adopts a specific supervised learning strategy. Traditional ANC algorithms (such as FxLMS) are constantly iterated during operation to minimise the total error energy. Unlike this, the Deep ANC system relies on the offline training stage to learn the nonlinear mapping from the reference signal to the optimal control signal. This section will detail the training framework based on physical paths, and explain the loss function derivation process and specific optimisation strategies specially designed for speech retention.

## 3.4.1 Supervised Learning Framework for Speech Preservation

In the training architecture of Deep ANC, we regard the CRN network as a nonlinear controller $f_\theta(\cdot)$, where $\theta$ represents the network parameters to be trained. The core goal of the training is to optimise $\theta$, so that the control signal $y(n)$ generated by the controller can effectively eliminate the primary noise after secondary path propagation.

In order to adapt the deep neural network to the physical characteristics of the ANC system, we have incorporated the acoustic response of the secondary path into the model training process. Referring to the method of Zhang and Wang [3], this study modelled the impulse response $s(n)$ of the secondary path as a fixed convolution layer, behind the output end of the CRN network.

Assuming that the network generates a predictive control signal $y(n)$ according to the input reference signal $x(n)$, then the actual anti-noise signal $a(n)$ of the arrival error microphone is defined as::

$$a(n) = y(n) * s(n) \tag{3.14}$$

Among them, $*$ represents the linear convolution operation. In the process of backpropagation, the secondary path parameter $s(n)$ remains frozen and does not participate in the update. However, the gradient information can still be transmitted back to the CRN controller through this layer. This design forces the neural network to automatically compensate for the amplitude and phase distortion caused by the secondary path while learning to generate anti-noise waveforms. Therefore, this method avoids the secondary path inverse modelling process commonly required in traditional methods [2].

Figure 3.6 shows the overall signal flow and gradient propagation path in the training stage. The solid line represents the forward transmission process of the signal generation, and the dotted line represents the reverse propagation process of the gradient through the fixed secondary path model.

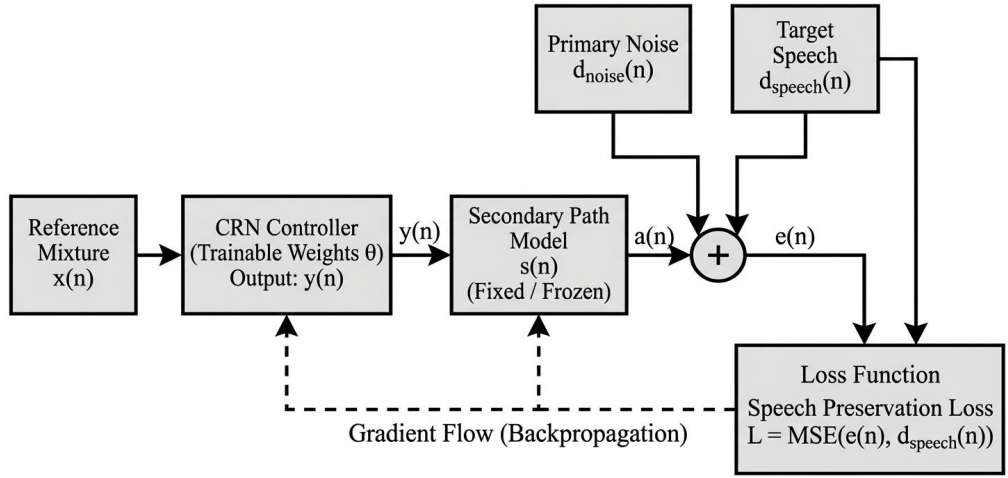

Figure 3.6 Signal flow diagram of the training process

## 3.4.2 Speech-Preserving Loss Function

The design of the loss function is the key to realising high-fidelity voice retention. The physical error signal $e(n)$ received by the error microphone is a linear superposition of the original interference signal and the anti-noise signal. In the scenario of this study, the original interference $d(n)$ consists of two parts: the environmental noise to be eliminated $d_{noise}(n)$ and the target voice to be retained

$d_{speech}(n)$:

$$d(n) = d_{noise}(n) + d_{speech}(n) \tag{3.15}$$

Therefore, the total signal $e(n)$ at the error microphone can be expressed as:

$$e(n) = d(n) + a(n) = [d_{noise} + d_{speech}(n)] + a(n) \tag{3.16}$$

where $a(n)$ is the anti-noise signal produced by the secondary loudspeaker, defined as the convolution of the control signal $y(n)$ with the secondary path impulse response $s(n)$ (see Eq. (3.14)).

In traditional ANC applications, the optimisation goal is usually to minimise the mean square error (MSE) of the error signal, that is, $\mathcal{C}_{trad} = E[e(n)^2]$. However, if this loss function is used directly, the network will tend to generate antitrust numbers to offset both noise and voice, resulting in a silent state. This obviously violates the original intention of retaining useful voices.

In order to achieve selective control, we introduced the concept of Acoustic Transparency. The ideal voice retention system should be transparent to the voice while eliminating noise, which means that the expected residual signal at the error microphone should converge to the pure target voice component $d_{speech}(n)$. Based on this, this study has developed a speech retention loss function $\mathcal{C}_{speech}$:

$$\mathcal{C}_{speech} = \frac{1}{L} \sum_{n=1}^{L} (e(n) - d_{speech}(n))^2 \tag{3.17}$$

Among them, $L$ is the length of the signal sequence. Suppose the expansion of $e(n)$ into the equation:

$$\mathcal{C}_{speech} = \frac{1}{L} \sum_{n=1}^{L} ([d_{noise}(n) + d_{speech}(n) + a(n)] - d_{speech}(n))^2 \tag{3.18}$$

After mathematical simplification, the phonetic components on the right side of the equation cancel each other. Finally, the loss function is simplified to:

$$\mathcal{L}_{\text{speech}} = \frac{1}{L}\sum_{n=1}^{L}(d_{\text{noise}}(n) + a(n))^2 \qquad (3.19)$$

The above derivation reveals the essence of the training strategy: by minimising the difference between the total error signal and the pure voice, the ultimate goal of minimising residual noise is indirectly achieved. This design builds a clever constraint mechanism. The network is trained to generate an anti-noise signal $a(n)$ that only approximates the inverse component of environmental noise (i.e. $a(n) \approx -d_{noise}(n)$), so as to realise phase cancellation interference to noise. At the same time, because the loss function does not contain speech residues, the network will not actively offset the voice component. This effectively realises the transparent transmission of voice signals [21].

It is worth noting that although the simplified formula (3.19) only involves noise, in actual training, the input of the network is always a mixed signal $x(n)$ containing voice and noise. This forces the Deep ANC model to learn to separate target speech features from complex mixed sound fields.

In summary, the Deep ANC system combines the CRN architecture with the supervised learning framework under physical constraints. By reconstructing the loss function based on the principle of acoustic transparency, this study theoretically solves the contradiction between noise cancellation and speech retention. Specific training details will be detailed in Chapter 4.

# Chapter 4
# Experimental Setup

This chapter details the experimental environment and methods for evaluating the Deep ANC system's performance. In order to ensure the reliability and reproducibility of the experimental results, we have built a rigourous test process. The content of the whole chapter is arranged as follows: Section 4.1 defines a multi-dimensional indicator system to evaluate the performance of the system, covering physical noise reduction effect and voice perception quality. And Section 4.2 describes the specific settings of the experiment, including the construction of the data set, the parameter configuration of the acoustic simulation environment, and the training details of the neural network. Section 4.3 introduces the FxLMS system as a comparison baseline and its configuration scheme under ideal conditions.

## 4.1 Performance Metrics

In order to comprehensively measure the performance of the speech-preserving Deep ANC system, this study builds a multi-dimensional evaluation system covering physical acoustics and perceptual acoustics. Our goal is not only to eliminate environmental noise at the physical level, but also to retain the target voice at the perceptual level. Therefore, the evaluation mainly focusses on the following two dimensions:

First, use Noise Reduction (NR) to quantify the system's ability to suppress noise.

Second, introduce speech perception quality (PESQ) and short-term objective comprehensionability (STOI). These metrics objectively assess the fidelity and clarity of the target speech signal.

### 4.1.1 Noise Reduction (NR)

To assess the physical efficacy of the ANC system, noise reduction (NR) is the most intuitive reference indicator. It reflects how much the noise energy at the error microphone position decreases before and after the control intervention.

In view of the fact that this study includes two scenarios of pure noise control and speech preservation, we have adopted a unified NR calculation definition to eliminate speech interference and focus on noise change:

1. Pure noise scenario: The error signal is completely composed of residual noise, and the energy ratio between the input noise and the output residual is directly calculated.

2. Speech preservation scene: In order to avoid the voice energy masking the noise reduction effect, the component separation strategy is adopted in the calculation. Only the energy change of the noise component in the error signal is counted, and the retained voice energy is not included.

Specifically, NR is defined as the logarithmic form of the energy ratio of the original environmental noise $d_{noise}(n)$ and the residual noise of the system output $e_{noise}(n)$. Its mathematical expression is as follows:

$$\text{NR} = 10\log_{10}\left(\frac{\sum_n d_{noise}^2(n)}{\sum_n e_{noise}^2(n)}\right) \qquad (4.1)$$

In the formula, $d_{noise}(n)$ represents the original noise signal that has not been suppressed, and $e_{noise}(n)$ represents the residual noise signal after the control takes effect. The numerical unit of NR is dB. The larger this value, the more obvious the suppression effect of the system on background noise.

## 4.1.2 Speech Quality and Intelligibility

In view of the system goal of retaining useful human voice while filtering out noise, it is impossible to fully evaluate the system performance by noise reduction alone. If the system introduces too much nonlinear distortion in the process of noise reduction, or deletes some audio segments by mistake, it will seriously affect the user's auditory experience. Therefore, this study introduces two industry standard indicators to evaluate the effectiveness of speech retention:

- **Speech Quality Perception Assessment (PESQ)**：

In order to evaluate the naturality and perceived quality of speech components in residual signals, we have adopted the PESQ algorithm defined by ITU-T P.862 standard [40]. The algorithm compares the difference between pure reference voice and enhanced voice by simulating the psychoacoustic characteristics of human hearing. The possible score ranges from -0.5 to 4.5. A higher score means the sound is more natural and the distortion is lower. In Deep ANC applications, a high PESQ score means that the controller successfully retains the voice signal and realises high-quality voice transmission.

- **Short-term objective understandability (STOI)**：

In order to evaluate whether the voice content is still clear and recognisable in the presence of residual noise and potential nonlinear distortion, we adopted the STOI indicator proposed by Taal et al. [41]. STOI is calculated based on the short-term envelope correlation of the time-frequency unit, which has been proved to be highly correlated with the word recognition rate in the artificial hearing test. Its value ranges from 0 to 1. The higher the score (tends to 1), the higher the clarity and comprehensibility of the voice. For ANC headphone applications involving human-computer interaction or communication, high STOI values are the key to ensure the accurate transmission of information.

These two indicators together constitute a comprehensive evaluation of the Deep ANC system, ensuring that it takes into account the fidelity of the signal while achieving the goal of noise reduction.

## 4.2 Experimental Settings

To test the effectiveness and robustness of the Deep ANC system in a complex acoustic environment, this study builds a high-fidelity simulation platform based on physical acoustics. The experimental settings cover many aspects such as data set construction, acoustic environment configuration, and neural network training details.

### 4.2.1 Datasets and Data Preparation

The dataset used in this experiment is a mixture of pure voice library and environmental noise library, aiming to simulate the real acoustic environment.

1. **Speech Dataset**

Clean speech data is taken from the test-clean subset within the LibriSpeech corpus [6]. The data set contains a large number of English audiobook fragments recorded by different readers, featuring high signal-to-noise ratio (SNR) and clearly structured speech. In order to match the typical working frequency of the ANC system, all voice data is re-sampled to 16 kHz.

2. **Noise Dataset**

Environmental noise data is selected from the NOISEX-92 database [5]. The training set covers the diversified noise types in the database to drive the deep neural network to fully learn the multi-type noise distribution law from steady-state low frequency to non-stationary broadband.

In the testing and evaluation stage, to quantitatively analyze the performance of the system from different dimensions of time-frequency characteristics, we specially selected the following five most representative noise sources as evaluation benchmarks:

- Engine: Low-frequency periodic engine noise, representing mechanical vibration interference.
- Factory1: Broadband non-stationary factory noise represents a complex industrial environment.
- Babble: The crowd is noisy, highly non-stationary, and the spectrum is highly overlapping with the voice, which is a very challenging source of interference.
- Volvo: Low-frequency steady-state noise in the car represents the transportation

scene.

- F16: Broadband jet noise represents high-intensity broadband interference.

**3. Data synthesis**

This study focuses on the two core tasks of pure noise reduction and speech preservation. For different task objectives, we use the Image Source Method (ISM) to build a unified acoustic simulation environment (room size 4m × 3m × 2.5m, reverberation time RT60=0.3s), and generate two sets of training data sets:

- For pure noise control tasks: Contains 10,000 samples. The reference signal $x(n)$ is generated by various noises in the NOISEX-92 library and the main path RIR convolution, including only ambient noise. The training target signal is set to a full zero vector, aiming to train the model to eliminate all incident sound waves.

- For speech preservation tasks: It also contains 10,000 samples, but adopts a dynamic mixing strategy. After mixing all kinds of ambient noise with LibriSpeech voice according to the random signal-to-noise ratio (0 0 dB to 10 dB), we convolution with the main path RIR respectively. The reference signal $x(n)$ is a mixture of speech and noise, and the training target $d_{target}(n)$ is defined as a pure speech component propagated through the main path.

During the test phase, we generated 100 fixed samples for each type for the above 5 representative noises (SNR = 5 dB). This is to quantitatively evaluate the final convergence effect and voice protection performance of the model in a specific acoustic scenario.

## 4.2.2 Acoustic Environment Simulation

Simple digital signal processing often fails to truly reflect the propagation characteristics of sound in physical space. In order to overcome this limitation, we used the *pyroomacoustics* toolkit to build a virtual acoustic laboratory [4]. Based on the Image Source Method (ISM) [29], the toolkit can accurately simulate multi-path propagation and reverberation effects in the room. The specific parameters of the simulation environment are as follows:

- Room size: Set to $4\text{m} \times 3\text{m} \times 2.5\text{m}$. This size simulates the interior space of a typical small and medium-sized office or vehicle.

- Reverberation time ($RT_{60}$) : Set to 0.3 seconds. This value represents an indoor environment with moderate acoustic treatment, which includes both early reflection sound and a certain degree of reverberation trailing. This tests the ANC system in terms of causality and convergence speed.

- Transducer Layout: Follow the physical constraints of single-channel feedforward ANC. The reference microphone is positioned near the noise source to capture the reference signal. Additionally, the distance between the secondary sound source and the error microphone is only 5 centimetres. This compact layout minimises the acoustic propagation delay, thus meeting the strict requirements of broadband noise control for causality.

Through the above settings, the system generates a room pulse response (RIR) for primary and secondary paths. Subsequently, these RIRs and synthetic data are convoluted to generate the final microphone signal.

## 4.2.3 Implementation Details

Deep ANC model is built on PyTorch deep learning framework and trained on NVIDIA GPU platform. In order to ensure the reprodusibility of the experiment and verify the feasibility of the system in real-time deployment, we have formulated a set of rigorous implementation plans, covering hyperparameter configuration, input signal processing and phased training strategies.

1. Input processing and frequency domain settings

The system operates at a 16 kHz sampling rate. In the spectrum feature extraction stage, we use 320-point FFT and 160-point Hop Length, which corresponds to 50% of the frame overlap rate. This configuration achieves the best balance between ensuring time-frequency resolution and reducing processing delay, and strictly meets the causal requirements of active noise control.

2. Hyperparameter configuration

The specific superparameters of model training are summarized in Table 4-1.

- Optimization strategy: Employ the Adam optimizer to handle the highly non-stationary dynamics of acoustic signals [39].

- Computational Efficiency: This study introduces Auto Mixed Precision (AMP) technology. The strategy dynamically mixes half-precision (FP16) and single-precision (FP32) floating-point values, which significantly reduces the memory footprint while maintaining numerical accuracy. In addition, we applied gradient cutting technology with a threshold of 5.0 in backpropagation to effectively prevent the gradient explosion of the LSTM layer.

| Category | Parameter | Value / Configuration |
|---|---|---|
| Acoustic Features | Sampling Rate | 16 kHz |
| | Frame Size ($N_{FFT}$) | 320 samples (20 ms) |
| | Hop Length | 160 samples (10 ms) |
| | Input Representation | STFT (Real, Imag) |
| Optimization | Optimizer | Adam |
| | Initial Learning Rate | 0.0005 |
| | Loss Function | Speech-Preserving MSE |
| | Gradient Clipping | Threshold = 5.0 |
| Training Setup | Batch Size | 16 |
| | Total Epochs | 70 (40 Pre-train + 30 Fine-tune) |

Table 4-1 Hyperparameters and Training Configurations

3. Two-Stage Training Strategy

In order to maximize the performance of the model and ensure convergence stability, we have implemented a training process that includes pre-training and fine-tuning:

- First stage (Pre-training, 40 Epochs): Set the initial learning rate to 0.0005. This stage aims to enable the network to quickly grasp the physical characteristics of the acoustic environment and the reverse mapping law of the secondary path.

- Second stage (Fine-tuning, 30 Epochs): On the basis of loading the optimal weight of the first stage, the learning rate is reduced to 0.00025 and 30 rounds of training are continued. This fine-tuning stage helps the model further converge near the minimal value of the loss function by updating more subtle parameters. Relying on the above optimization strategy, the Deep ANC system shows efficient convergence ability and robust performance with limited computing resources.

## 4.3 Baseline System

In order to objectively verify the performance advantages of Deep ANC system in controlling nonlinear and broadband noise, we chose the classic filter-x minimum mean square (FxLMS) algorithm as the comparison baseline. With its simple structure and extremely high computing efficiency, the FxLMS algorithm has long been regarded as the industry standard in the field of active noise control [2, 7].

### 4.3.1 Principle of the FxLMS Algorithm

FxLMS algorithm is essentially a linear adaptive feedforward control system. It uses the adaptive filter $W(z)$ to generate anti-noise signals, aiming to minimise the mean square of the error signal $e(n)$. To counteract the phase delay and amplitude distortion induced by the secondary path $S(z)$, the algorithm filters the reference signal $x(n)$. . The corresponding weight update rule is given by:

$$w(n+1) = w(n) - \mu \cdot e(n) \cdot x'(n) \quad (4.2)$$

Among them, $w(n)$ is the weight vector of the control filter, $\mu$ is the step parameter, and $e(n)$ represents the error signal. $x'(n)$ is a reference signal filtered by the estimated secondary path model $\widehat{S}(z)$:

$$x'(n) = \widehat{S}(n) * x(n) \quad (4.3)$$

This mechanism ensures the correctness of the gradient descent direction, thus ensuring the stability of the system [2].

## 4.3.2 Configuration and Implementation

In order to ensure comparative rigor and fairness, we evaluate the FxLMS baseline system in an acoustic simulation environment that is completely consistent with Deep ANC (see Section 4.2.2 for details). Based on the actual operation logic of Python simulation code, the specific parameter configuration and implementation details are as follows:

- Control Filter: The controller uses a finite impulse response (FIR) filter, and its length (Order) is set to 512. This length is enough to cover the main acoustic delay of the secondary path, thus meeting the strict requirements of broadband noise control for the causality of the system.

- Step Parameter: The learning rate $\mu$ is set to 0.001. After experimental verification, the value has achieved an effective balance between the convergence speed and the steady-state error.

- Ideal Secondary Path Estimation: In order to build a strong baseline and eliminate the interference of estimation errors on FxLMS performance, this study assumes that the secondary path information is completely known. In the code implementation, we directly use the real RIR generated by the simulation as the estimate value $\widehat{S}(n)$ in the algorithm to calculate the filter reference signal $x'(n)$.

- Testing Protocol: The FxLMS algorithm runs in online Sample-by-sample mode.

In order to ensure consistency with Deep ANC test data, we set up a fixed random seed (Fixed Seed = 42) and evaluated it on the same 100 test samples. When calculating performance indicators (NR), we remove the data of the first 4000 sampling points (about 0.25 seconds) of each sample to exclude transient fluctuations in the initial convergence stage of the algorithm, and only count the performance of the system after it tends to stabilize.

# Chapter 5
# Results and Analysis

This chapter aims to comprehensively evaluate the performance of speech-preserving Deep ANC systems in a high-fuly simulation reverberation environment. To confirm that the system outperforms conventional methods, we have made an in-depth analysis from the two dimensions of physical noise reduction effect and speech perception quality. The content of the whole chapter is arranged as follows. Section 5.1 first establishes a physical noise reduction benchmark in pure noise scenarios and makes a quantitative comparison with the traditional FxLMS algorithm. Section 5.2 focusses on the core function of the system - selective noise cancellation, and evaluates its ability to retain the target voice in the mixed sound field through qualitative and quantitative indicators. Next, Section 5.3 makes an in-depth analysis from the mechanical level, revealing the physical reasons why Deep ANC surpasses the baseline system in terms of transient response and nonlinear processin. Finally, Section 5.4 verifies the generalisation ability and robustness of the model in the face of noise types and complex acoustic characteristics.

## 5.1 Noise Reduction Performance in Pure Noise Scenarios

This section aims to explore the limit of physical noise reduction of Deep ANC system in a pure noise-free environment without voice interference. This is a key link in evaluating the benchmark performance of the active noise control system. We put the system in a simulated real reverb environment ($RT_{60} = 0.3s$) and selected five noise sources with different time-frequency characteristics for testing.

### 5.1.1 Quantitative Analysis

In order to objectively quantify the noise reduction capacity of the system, we use noise reduction (NR) as the core indicator. To validate the effectiveness of the new method, we compared Deep ANC with the classic linear FxLMS algorithm. The specific experimental results are listed in Table 5-1, and Figure 5.1 intuitively shows the performance differences between the two.

| Noise Type | Category | FxLMS (dB)* | Deep ANC (dB) | Improvement |
|---|---|---|---|---|
| **Engine** | Periodic | 12.22 | 22.92 | **+10.70 dB** |
| **Babble** | Non-stationary | 5.28 | 18.17 | **+12.89 dB** |
| **Factory1** | Non-stationary | 6.17 | 18.49 | **+12.32 dB** |
| **Volvo** | Stationary | 4.91 | 19.08 | **+14.17 dB** |
| **F16** | Broadband | 8.71 | 17.35 | **+8.64 dB** |

Table 5-1 Performance comparison between FxLMS and Deep ANC in Pure Noise

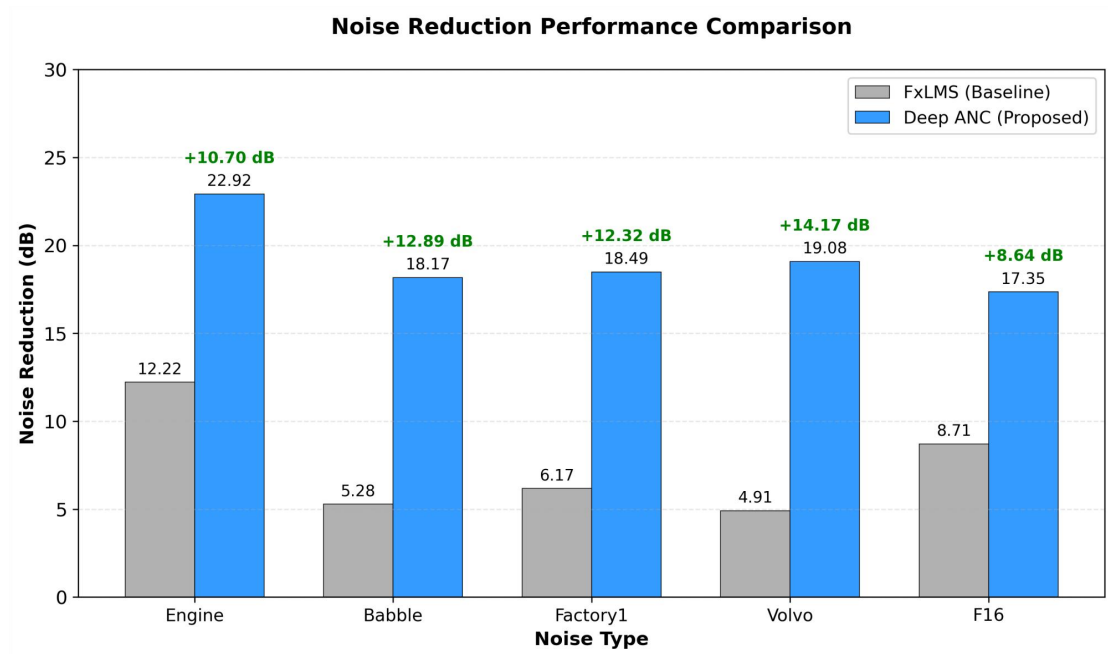

Figure 5.1 Noise Reduction Performance Comparison

Looking at the experimental data in Figure 5.1 and Table 5-1, the Deep ANC system showed significantly better performance than traditional methods in all test scenarios. Through in-depth analysis of the performance of different noise types, we can draw the following three key conclusions:

**1. Deep suppression for periodic and steady-state noise**

In scenarios with significant periodic or steady-state characteristics such as Engine and Volvo, Deep ANC achieves the highest noise reduction, reaching 22.92 dB and 19.08 dB respectively. Although FxLMS can also achieve 12.22 dB in the Engine scenario (which is the best field of linear algorithms), Deep ANC still achieves a huge increase of +10.70 dB. This proves that the CRN network has a strong timing correlation modeling ability, which can predict and offset low-frequency harmonic components more accurately than linear adaptive filters.

**2. Break through the control bottleneck of non-stationary noise**

The most significant performance gap occurs in non-stationary noise scenarios such as Babble and Factory1. For Babble, a rapidly changing noise full of random burst signals, the traditional FxLMS algorithm is difficult to track the sharp fluctuations of its statistical characteristics, and the noise reduction is only 5.28 dB. In contrast, Deep ANC achieves 18.17 dB of noise reduction in this scenario, which is +12.89 dB higher than the baseline. This result strongly proves that the LSTM unit successfully captures the long-time sequence dependencies in non-stationary signals, so that the system can maintain efficient prediction and suppression in the face of complex and variable acoustic interference.

**3. Full band coverage under wideband noise**

In the F16 (jet) broadband noise scenario, Deep ANC also maintains a high level of 17.35 dB. This shows that the system is not only effective in the low frequency band, but also its convolution structure can extract effective acoustic characteristics in a wide frequency band range. In contrast, traditional algorithms are limited by the linear hypothesis and convergence speed of filters, and are often unable to handle broadband signals.

In summary, the Deep ANC system overcomes the limitations of traditional linear algorithms in non-stationary and broadband noise processing. Whether in a regular mechanical noise environment or chaotic crowd interference, the model can maintain a stable noise reduction level of more than 17 dB, showing strong environmental adaptability and robustness.

## 5.1.2 Physical Characteristics and Real-time Feasibility

To gain deeper insight into the system's dynamic response mechanism, we selected two very representative noises for micro waveform analysis:

- Engine: represents stationary periodic noise, characterized by significant harmonic structures and stable frequency distribution.

- Babble: represents non-stationary noise, featuring highly unstable, chaotic, and transient fluctuations caused by overlapping speech.

Figure 5.2 shows the comparison between the time domain waveform of these two noises before and after control (the figure above) and the spectrum diagram (the figure below).

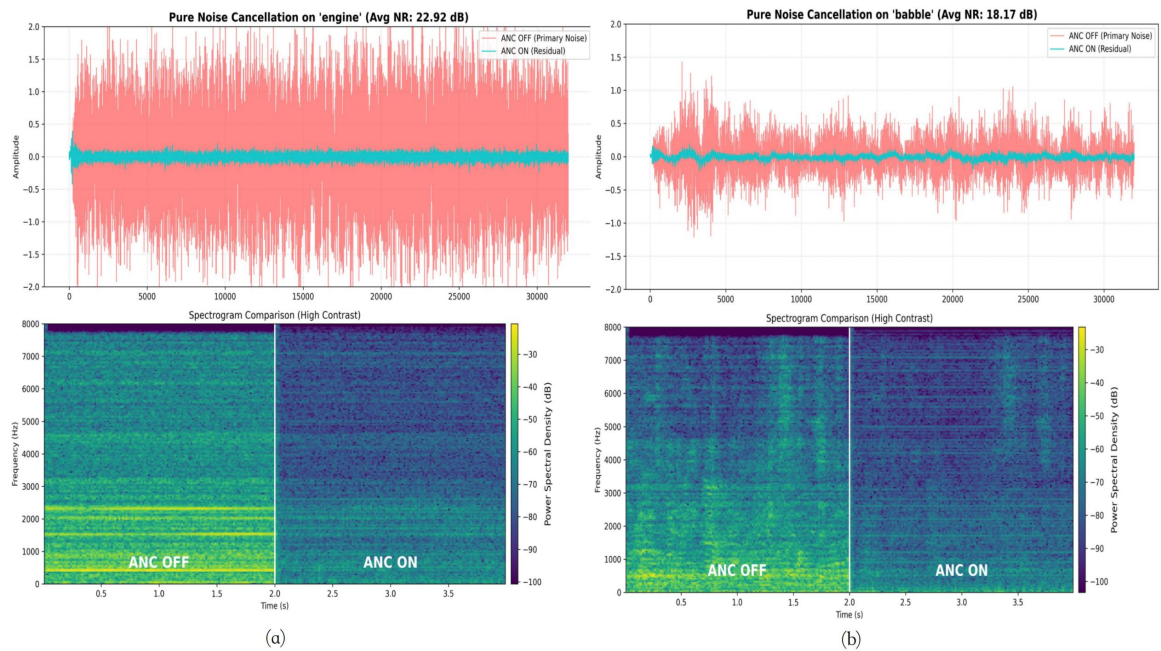

Figure 5.2 Comparison of time-domain waveforms (top) and spectrograms (bottom) before and after ANC control

*Note: (a) Periodic Engine noise, (b) Non-stationary Babble noise.*

1. **Precise harmonic elimination of periodic noise (Engine Case)**

As shown in Figure 5.2(a), the Engine scene shows the ultimate control ability of the system for steady state noise.

- Time domain analysis: The original noise (red waveform) shows a large periodic oscillation. After the system intervenes, the residual error signal (cyan waveform) is compressed to a flat line close to the zero axis, achieving energy attenuation of up to 22.92 dB.

- Time-frequency domain analysis: The spectrogram clearly reveals the frequency structure of the noise. In the ANC OFF area, distinct bright yellow and green horizontal bands can be observed, which are low-frequency harmonics generated by mechanical operation. In the ANC ON area, these high-energy yellow bands are completely erased, and the background turns into a uniform deep purple. This proves that the CRN convolution filter successfully extracts the spectrum texture with translation invariance.

2. **Dynamic tracking of non-stationary noise (Babble Case)**

Figure 5.2 (b) shows the most challenging Babble scene. Unlike Engine, Babble noise consists of a large number of overlapping vocal fragments and lacks predictable periodicity.

- Roust suppression: Although the input signal is full of chaos, the system still achieves a significant noise reduction of 18.17 dB. In the time domain, the random sudden peak in the red waveform is effectively clamped by the cyan waveform.

- Spectrum smoothing: In the spectrum diagram, the original signal (ANC OFF) presents scattered bright yellow patches representing high-energy speech resonance peaks. After treatment (ANC ON), these high-energy highlights are effectively suppressed, and the overall tone of the map changes from warm (yellow and green) to cold (dark blue and purple). This shows that although there are still trace textures affected by the non-steady characteristics, the high-intensity energy peak has been completely eliminated. The results confirm that the LSTM bottleneck layer effectively captures the time correlation of non-steady signals.

3. **Frequency domain harmonic comb filter effect**

In order to quantify the harmonic elimination effect from the perspective of the frequency domain, we further draw a power spectral density (PSD) comparison chart of Engine noise, as shown in Figure 5.3.

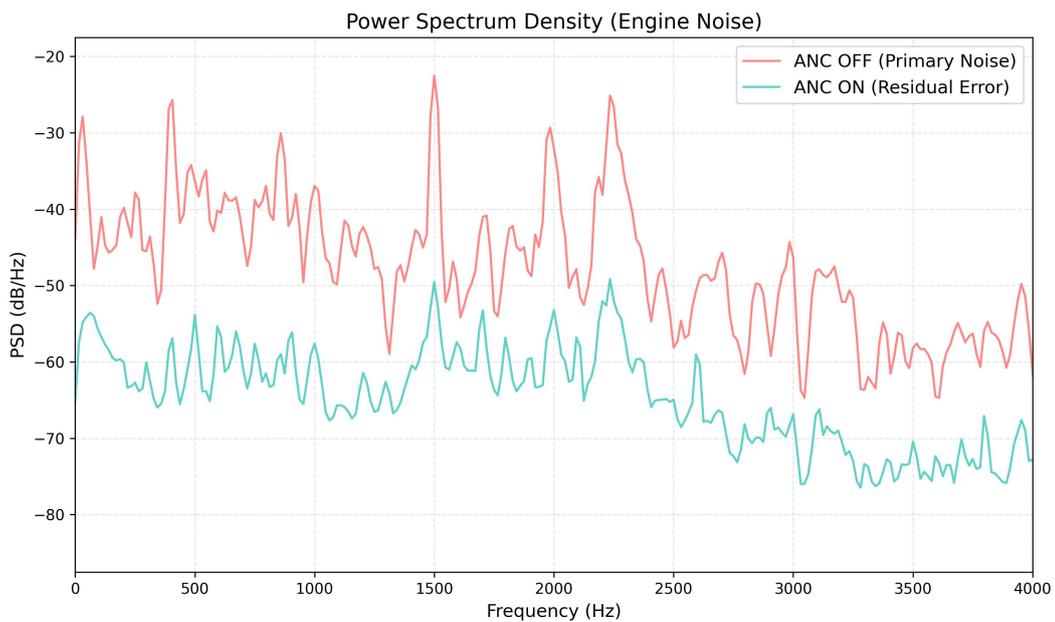

Figure 5.3 Comparison of Power Spectral Density (PSD) for Engine noise

Figure 5.3 clearly shows the harmonic comb filter characteristics of the Deep ANC system.

- ANC OFF (red line): There are multiple sharp energy peaks at the base frequency and its harmonic frequencies (especially around 1500Hz and 2200Hz).
- ANC ON (cyan line): These spikes are accurately leveled, and the attenuation of local frequency points exceeds 20 dB.

This result once again confirms that the Deep ANC system does not blindly reduce the full-band energy, but can accurately identify and strike specific frequency points of noise energy concentration. At the same time, there is no abnormal clutter component in the cyan line, indicating that the system does not introduce additional nonlinear distortion while realising deep noise reduction.

It is worth mentioning that the CRN architecture of this system strictly follows the principle of Causal Convolution in its design. This means that when producing the anti-noise signal for the current frame, the network only relies on the current and past historical information, and never uses future frames. This design theoretically ensures the physical realisation of the algorithm and lays the foundation for future porting to low-latency hardware platforms such as FPGA or DSP.

## 5.2 Selective Noise Cancellation and Speech Preservation

In the previous section, we explored the physical noise reduction limits of the Deep ANC system in a pure noise environment. However, real application scenarios are often more complex. The user's environment is usually a mixed sound field, that is, the target voice is superimposed with environmental noise. Due to the lack of capability to distinguish speech from noise, the traditional FxLMS algorithm often inevitably damages useful voice signals while eliminating broadband noise, resulting in a decline in speech quality.

This section aims to deeply verify whether the CRN architecture has true spectral selectivity. We need to examine whether this system can use the powerful nonlinear feature extraction ability of the deep neural network to accurately identify and separate the target voice in the mixed signal with highly overlapping acoustic features, so as to achieve the goal of denoising and sound preservation.

To simulate a realistic mixed sound field, we dynamically superimpose clean speech from the LibriSpeech dataset (as the target signal) with ambient noise from the NOISEX-92 dataset. During training, the signal-to-noise ratio (SNR) for mixing is randomly chosen between 0 dB and 10 dB. This helps the model adapt to different noise levels. In contrast, for the testing phase, the SNR is strictly fixed at 5 dB to ensure a standardized and representative evaluation benchmark.

## 5.2.2 Visual Analysis: Spectrograms

In order to intuitively verify the characteristic decoupling ability of the Deep ANC system in the time-frequency domain from a micro perspective, we selected two highly representative mixed sound fields for comparative analysis:

1. Engine noise: Represents steady-state interference with a strong periodic harmonic structure.

2. Babble (Crowd Noisy): Non-stationary vocal interference that represents highly overlapping and irregular spectrum

Figure 5.5 shows the reshaping effect of the system on spectrum characteristics in the Engine noise environment.

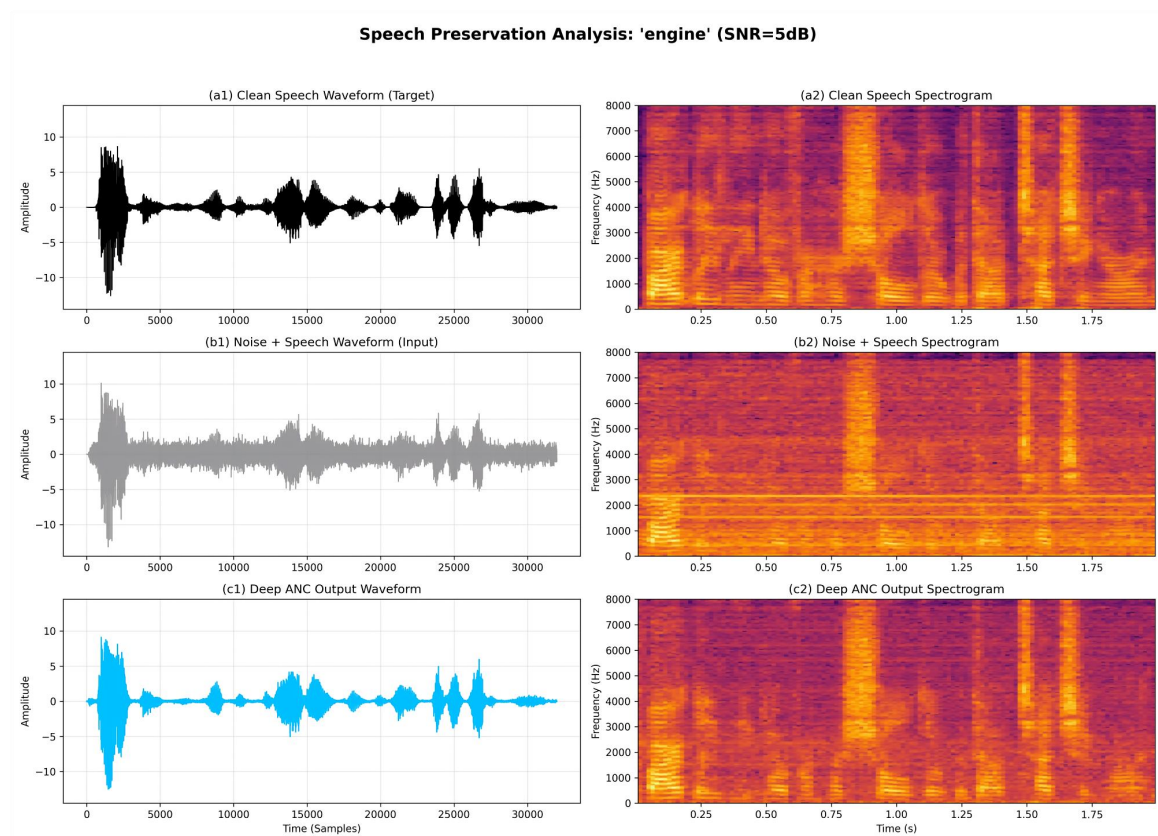

Figure 5.5 Spectrogram analysis of harmonic noise cancellation (Engine scenario)

*Note: Top: (a) Clean Speech, (b) Noise + Speech input (Obvious horizontal harmonics can be seen), (c) Deep ANC Output（The harmonics are eliminated and the voice resonance peak is retained）*

**1. Accurate harmonic stripping of periodic noise (Engine Case)**

From the input spectrum diagram of Figure 5.5 (b2), it can be clearly observed that there are many bright yellow horizontal lines with extremely high energy in the low-frequency area below 2000Hz, which is the mechanical harmonic generated by the engine operation. These horizontal textures seriously obscure the details of the voice.

In the processed figure 5.5 (c2), Deep ANC shows extremely high accuracy:

- Horizontal denoising: Those high-energy horizontal horizontal lines are completely erased, and the background area is restored to a deep dark purple, indicating that the steady-state noise energy is completely suppressed.

- Vertical fidelity: More importantly, the longitudinal resonance peaks (Formants) representing phonetic characteristics (bright columns distributed vertically in the figure) are not only not damaged, but also become clearer and sharper due to the removal of background interference. Such a visual comparison strongly proves that the CRN encoder has successfully learned the decoupling representation of orthogonal features.

Next, we focus on the most challenging Babble scenario, whose spectrum analysis is shown in Figure 5.6.

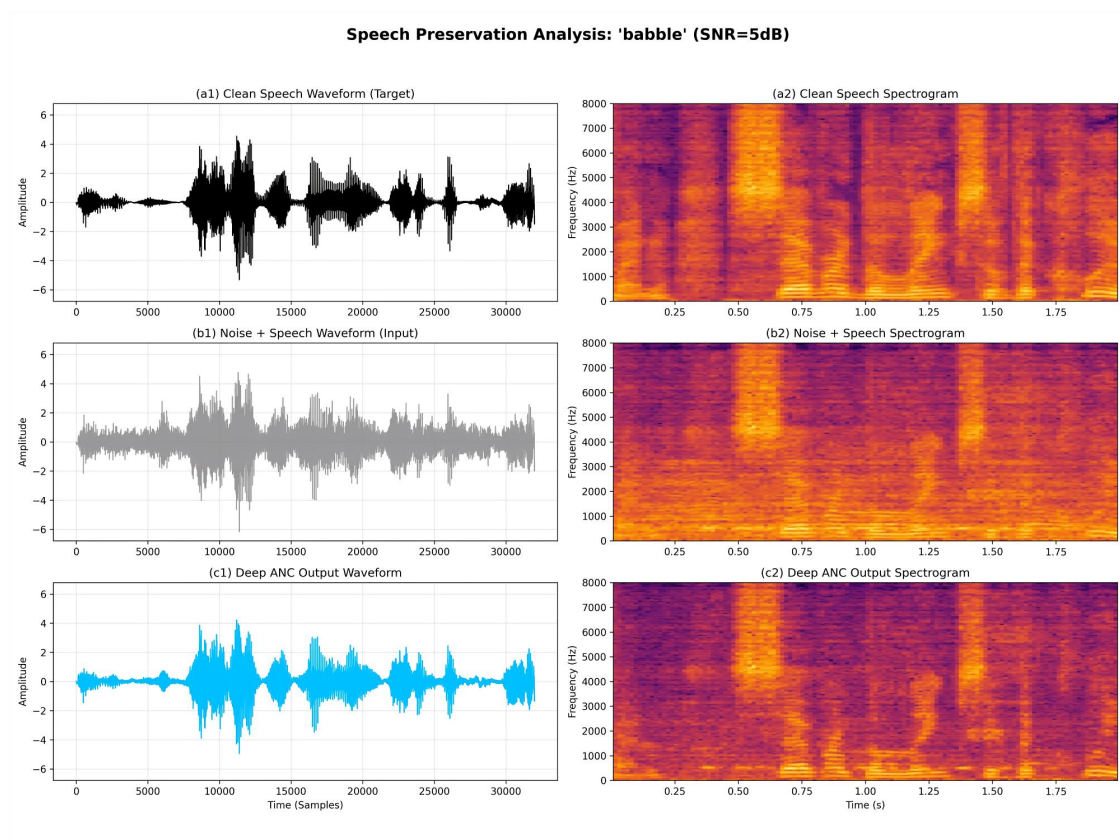

Figure 5.6 Spectrogram analysis of semantic speech separation (Babble scenario)

*Note: Top: (a) Clean Speech, (b) Noise + Speech input (Background Chaos), (c) Deep ANC Output (The background is blurrified, and the voice subject is prominent)*

## 2. Selective separation of non-stationary human voices (Babble Case)

Unlike Engine, Babble noise (Figure 5.6 b2) presents a haze-like chaotic texture in the full frequency band. Because the background itself is also a human voice, its spectrum and the target voice overlap highly in the time-frequency domain, so the traditional filter can't help it.

- Background blur: Looking at Figure 5.6 (c2), the most significant change is that the overall tone changes from bright orange to darker dark purple, which means that the energy density of the background voice is greatly weakened.

- Semantic locking: Although the background is suppressed, the skeleton structure (vertical bright column) of the target voice still stands upright and the edges become clearer. The system does not blindly erase all sounds, but intelligently lowers the background volume, achieving a background blur effect similar to that in photography. This confirms that the LSTM unit can accurately lock and track the semantic characteristics of the target speaker from chaotic scenarios based on long-time sequential context information.

Through this microscopic comparison, we have confirmed that the Deep ANC system can not only handle simple mechanical noise, but also has the advanced ability to separate target speech signals in complex mixed sound fields.

## 5.2.1 Quantitative Analysis

In order to quantify the effect of speech retention, we have introduced two core indicators: PESQ is used to evaluate the naturalness of hearing, and STOI is used to evaluate the clarity of content.

To build broad adaptability, the model is trained under a varying signal-to-noise ratio between 0 dB and 10 dB. In the test phase, in order to make a strict evaluation on a unified and representative benchmark, we fixed the signal-to-noise ratio of all test samples to 5 dB. This intensity represents a typical medium interference environment (such as a noisy office or a moving carriage), which can not only reflect the destructive power of noise, but also conform to the common scenes of actual calls.

Table 5-2 shows the comprehensive processing performance of the system for five

different noise types at a 5 dB signal-to-noise ratio.

| Background Noise | Category | NR (dB) | PESQ Improvement | STOI Improvement |
|---|---|---|---|---|
| Engine | Periodic | 8.88 | +0.686 | +0.101 |
| Babble | Non-stationary | 5.30 | +0.359 | +0.066 |
| Factory1 | Non-stationary | 7.22 | +0.444 | +0.080 |
| Volvo | Stationary | 13.70 | +0.464 | +0.001 |
| F16 | Broadband | 8.19 | +0.632 | +0.100 |

Table 5-2 Speech Preservation Performance of Deep ANC in Mixed Acoustic Scenarios (SNR = 5 dB)

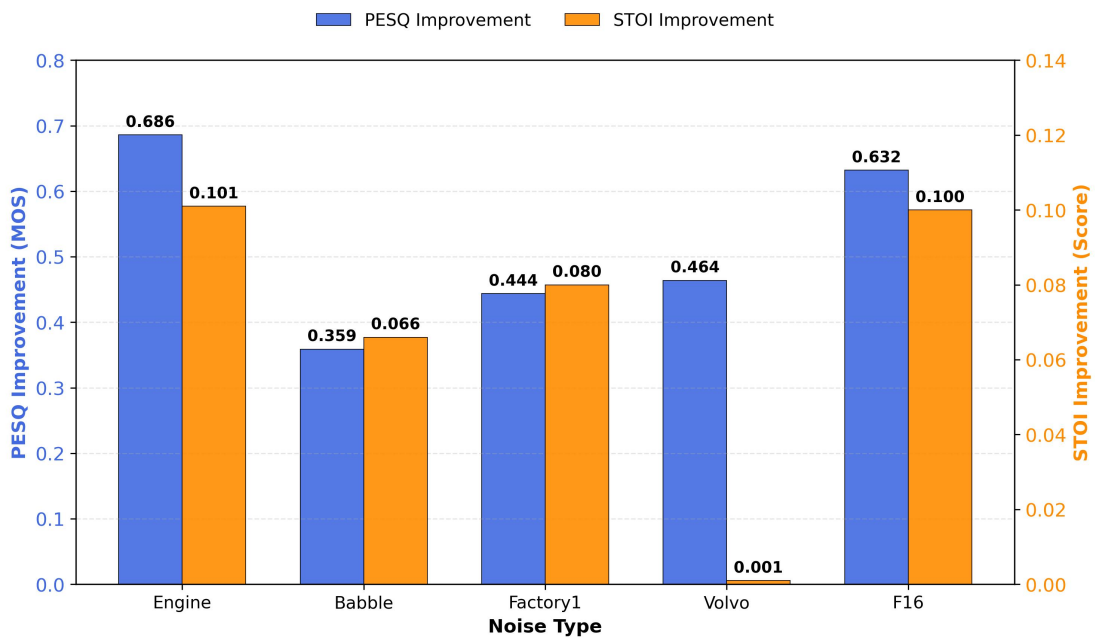

Figure 5.4 Improvements in Speech Quality (PESQ) and Intelligibility (STOI) across Different Noise Types

Combined with the data in Table 5-2 and the trend of Figure 5.4, and by analyzing the physical properties of different noises, we can summarize three key characteristics of Deep ANC in mixed sound field processing:

1. **High-fidelity reduction of periodic and broadband noise**

In the Engine and F16 scenarios, the system shows the most significant improvement in voice quality, with PESQ increments as high as +0.686 and +0.632 respectively. This shows that the CRN architecture can efficiently strip mechanical noise (Engine) and broadband high-frequency noise (F16) with a stable harmonic structure from the mixed signal, and greatly restore the clarity and natural sense of the voice while physical noise reduction (about 8-9 dB).

2. **Deep suppression of low-frequency steady-state noise**

In the Volvo scenario, the system recorded the highest noise reduction of 13.70 dB. Since this kind of noise energy is mainly concentrated in the low-frequency steady-state area, this is the frequency band that the ANC system is best at processing. Although the improvement of STOI is not obvious (+0.001, because the original voice itself is more understandable), the PESQ improvement of up to +0.464 proves that the increase in background quietness significantly improves the subjective listening experience.

3. **Intelligent balance of non-stationary vocal interference**

The most noteworthy is the Babble scene. This is a very challenging kind of interference, because the background noise and the target voice overlap highly on the spectrum. The experimental results show that although the system controls the physical noise reduction at a conservative level of 5.30 dB in order to protect the target voice from accidental injury, the PESQ is still increased by +0.359 and the STOI is increased by +0.066. This shows that the system does not blindly pursue high NR, but intelligently suppresses the most disruptive noise in the background, and successfully finds the best balance between noise reduction and voice fidelity.

### 5.2.3 High-SNR Waveform Fidelity Analysis

In the quantitative analysis above, we observed a significant phenomenon: in the Volvo scenario, although PESQ has been greatly improved (+0.464), the increase in STOI is negligible (+0.001). In order to explore the physical reasons behind this phenomenon and verify whether the system will cause damage to the voice signal itself in a low-noise interference environment, we have drawn a sample-level time-domain waveform comparison chart (Figure 5.7).

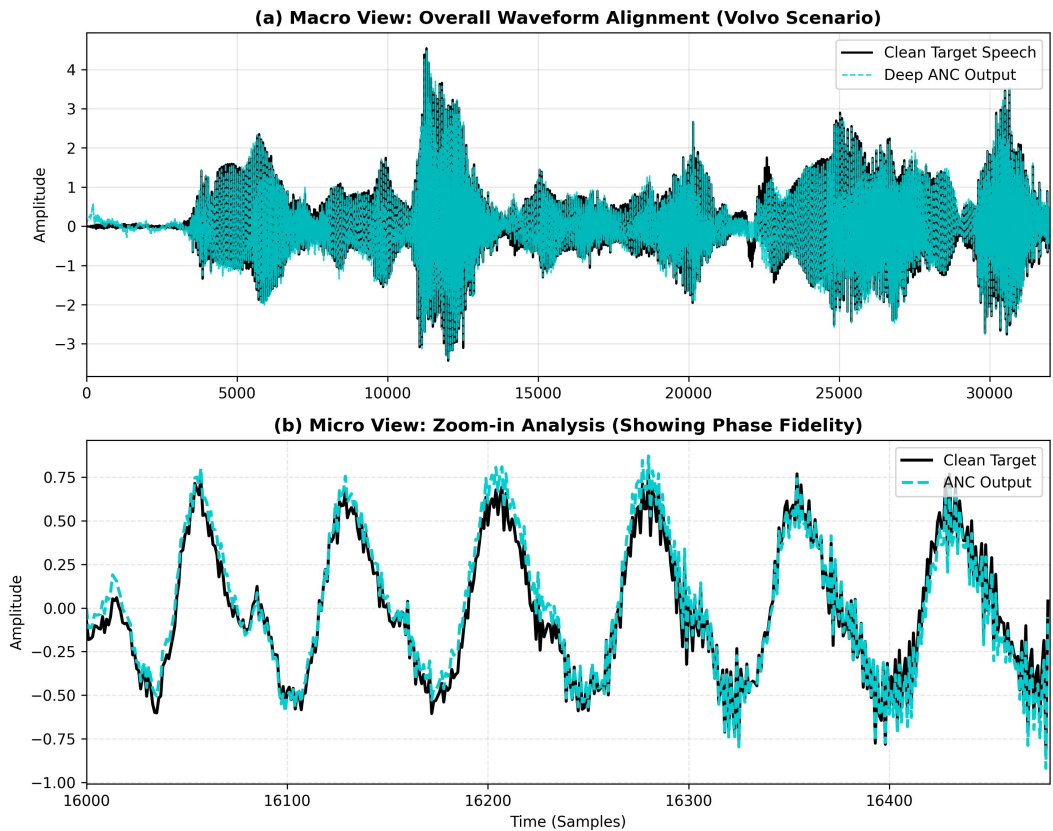

Figure 5.7 Waveform fidelity analysis in Volvo scene

*Note: (a) shows the perfect overlap of the overall envelope, (b) shows the precise alignment of the phase.*

Through the in-depth comparison of macro panorama and micro details, we can draw the following conclusions:

**1. The height overlap of the overall waveform**

As shown in Figure 5.7 (a), in the full time period of 2 seconds, the waveform output by Deep ANC (cyan dotted line) almost completely coincides with the original pure voice (black solid line) in shape. Especially in areas with large voice and sound (near the sampling point 12000), the ups and downs of the two are exactly the same. This shows that the energy envelope structure of the voice signal is completely retained when the system is noise reduction processing. This also explains why the STOI score has not decreased, because the main skeleton of the voice is intact.

**2. Phase alignment at the detail level**

In order to see the details, we zoomed in and observed the 30 millisecond fragment, as shown in Figure 5.7 (b).

- Phase accuracy: The peaks and valleys of the cyan waveform are accurately aligned with the position of the black waveform, and there is no left-right offset. This proves that the system processes the signal very accurately and does not change the phase information of the sound.

- Characteristics of residual noise: The cyan waveform has a slight jagged fluctuation on the smooth black baseline. These fluctuations are not caused by speech distortion, but by low-frequency residual noise that has not been completely eliminated. This shows that the system tries its best to reduce the noise on the basis of retaining the voice.

Based on the above analysis, the data characteristics (PESQ enhancement, STOI stability) in the Volvo scenario reflect the extremely high signal fidelity of the Deep ANC system. With the good quality of the input signal itself, the system can accurately identify and filter out the low-frequency steady-state noise in the background, while strictly protecting the voice component from accidental injury. This proves that the algorithm is safe and reliable in practical application and will not destroy the original clear voice due to noise reduction processing.

## 5.3 Comprehensive Comparative Analysis

In the previous experiment, the Deep ANC system showed significantly better performance than traditional methods. In order to explore the root cause of this difference, this section will deeply analyze the mechanical differences between the two algorithms from the physical acoustic level. We will focus on comparing their transient response characteristics under non-stationary signals and the modeling ability of nonlinear harmonics. In addition, this section will also verify the robustness and adaptability of the system in a complex and variable acoustic environment through macro-wide scenario evaluation.

### 5.3.1 Transient Response in Non-stationary Environments

Next, we further analyze the deep mechanism of Deep ANC's performance jump compared with the traditional FxLMS algorithm from the microscopic level of physical acoustic response. This performance gap is mainly due to the essential difference between the transient response characteristics and nonlinear modeling capabilities of the two algorithms.

**1. Advantages of transient response in a non-stationary environment**

The FxLMS algorithm is essentially an adaptive linear filter based on gradient descent, and its weight update depends on the stability of the statistical characteristics of the input signal. In the face of highly non-stationary signals such as Babble, its spectral characteristics fluctuate sharply over time. The convergence speed of FxLMS often lags behind the change speed of noise, resulting in the failure of the filter to reach a steady state.

Unlike traditional methods, with the help of LSTM's long-term memory mechanism, Deep ANC can instantly generate reverse phase waves according to the current frame characteristics, achieving a millisecond-level fast response.

Figure 5.8 intuitively shows the time-domain residual comparison of the two algorithms when dealing with Babble noise.

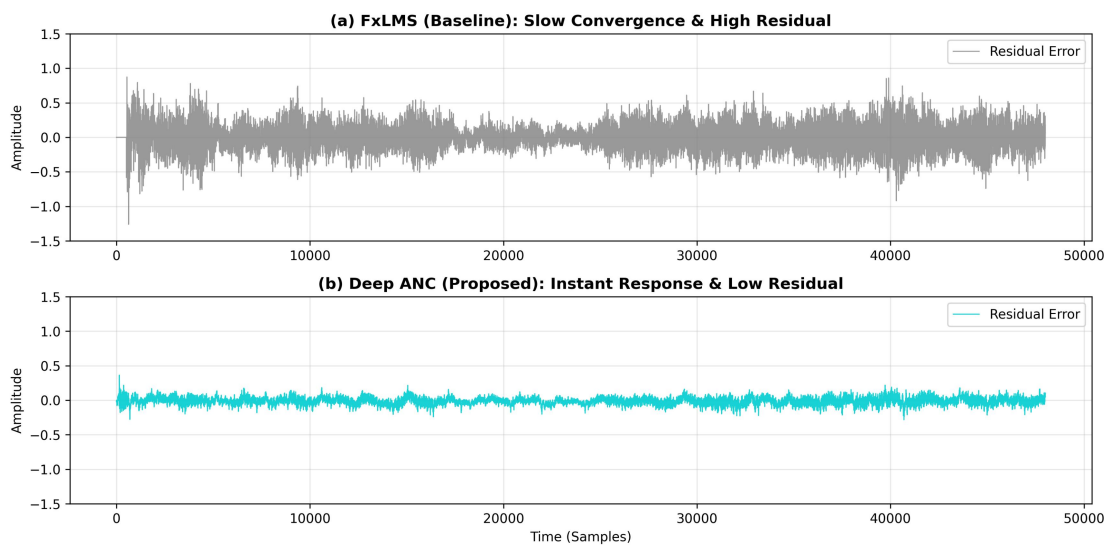

Figure 5.8 Comparison of transient response characteristics under Babble

- FxLMS (gray): The residual signal always maintains a high amplitude of oscillation, indicating that the system has been in an underconvergent state and cannot effectively track the sudden changes of the human voice.

- Deep ANC (cyan): The system enters steady-state control at the moment of startup, and the error signal is clamped at a very low level.

**2. The depth suppression ability of nonlinear harmonics**

In addition to non-stationaryness, nonlinear effects in physical paths (such as speaker harmonic distortion) are also bottlenecks that limit the performance of traditional ANC. Based on the linear hypothesis, FxLMS can only offset the base frequency component, and can't do nothing about the nonlinear harmonic beam.

In order to quantify this difference, we compared the residual power spectral density (PSD) processed by two algorithms in the Engine scenario, as shown in Figure 5.9.

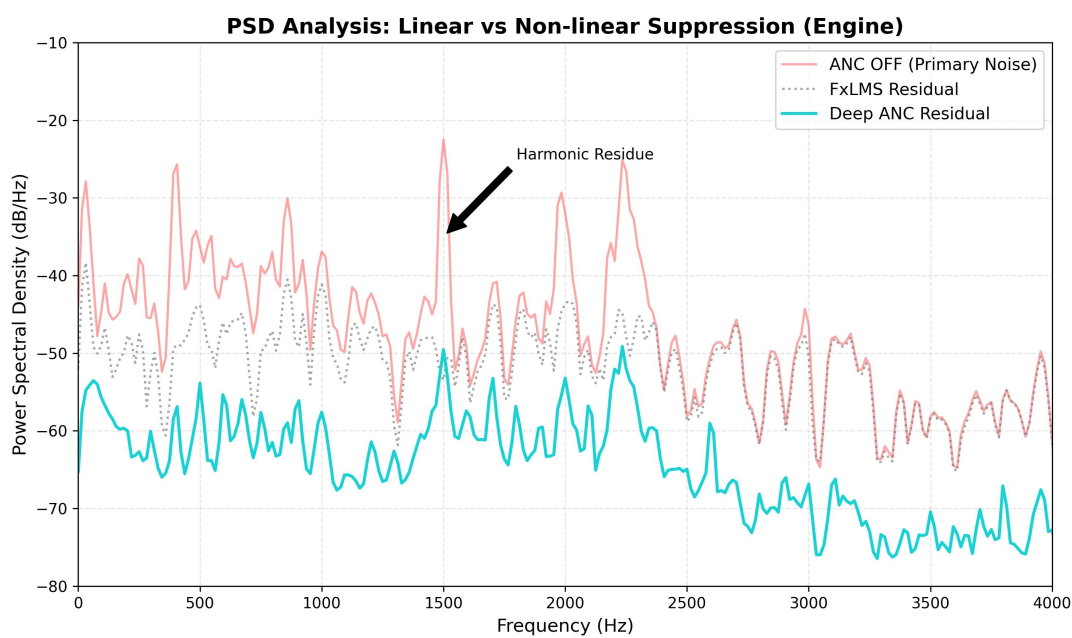

Figure 5.9 Power Spectral Density (PSD) analysis of linear vs. non-linear harmonic suppression (Engine scenario)

By looking at the three curves in the figure, we can clearly see:

- Original noise (red line): Around 1500Hz and 2200Hz, there are sharp peaks with extremely high energy. These peaks are the complex harmonic components generated by engine operation.

- FxLMS (grey dotted line): Although the gray line is a little lower than the red line as a whole, indicating that it plays a certain role in noise reduction, but at those key harmonic frequency points (such as 1500Hz pointed by the arrow), the gray line still retains an obvious spike. This shows that although the traditional linear filter can reduce the overall noise, it cannot eliminate these complex nonlinear harmonic residues.

- Deep ANC (cyan solid line): The most significant change is that the peaks of the green line at 1500Hz and 2200Hz have been completely flattened. At these frequency points, Deep ANC has 10 to 15 decibels less energy than FxLMS.

The Deep ANC system not only makes the sound smaller, but also uses the powerful nonlinear mapping ability of the deep neural network to accurately predict and offset the complex harmonic distortion of those traditional algorithms, and achieve real depth noise reduction [50].

## 5.3.2 Robustness to Diverse Noise Types

In the previous mechanism analysis, we have verified the microscopic advantages of the Deep ANC system in transient response and nonlinear processing. In order to further verify the generalization potential of the system at the macro level, this section comprehensively evaluates the performance distribution of the model in five typical noise scenarios with completely different time-frequency characteristics. Figure 5.10 intuitively shows the wide-spectrum adaptability of the system to various types of acoustic interference in the form of radar diagrams.

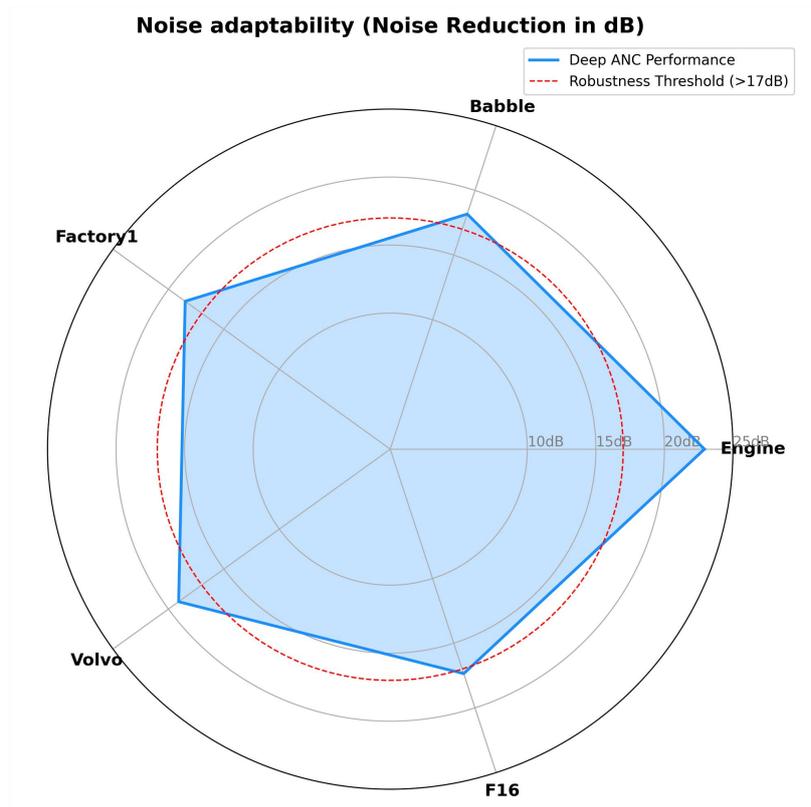

Figure 5.10 Radar chart of noise reduction performance

By analyzing the coverage area of the radar map, we can draw the following conclusions:

1. **Cross-spectrum Adaptability**

As shown in the figure, the performance envelope of the system on all five test dimensions significantly exceeds the robustness threshold (red line) of 17 dB. Whether the energy is mainly concentrated in low-frequency steady-state noise (represented by Volvo, 19.08 dB) or broadband high-frequency noise with a wide energy distribution (represented by F16, 17.35 dB), the model can maintain a high level of noise reduction. This proves that the CRN architecture does not rely on specific frequency distribution characteristics, but successfully learns the general law of acoustic signal processing, which can effectively cover various noise types from narrow to broadband.

**2. Resistance to Non-stationary Interference**

In the face of chaotic non-stationary scenarios where traditional linear algorithms are very easy to fail (such as Factory1 and Babble), Deep ANC still maintains a stable performance of more than 18 dB. This kind of noise has strong randomness and time variation, but the experimental results show that the model successfully overcomes the control lag caused by the rapid fluctuation of the input signal with the time memory mechanism of the LSTM unit. This confirms that the system can not only handle regular mechanical sound, but also calmly cope with complex acoustic interference caused by human voice superposition or sudden impact.

**3. Unified Processing Capability**

Traditional algorithms often need to adjust the control parameters for different noise characteristics (such as stationary or non-stationary) to maintain stability. In contrast, the system can automatically adapt to all the above types of interference without any manual intervention or parameter switching with only a single pre-training model. This ability to deal with diversified noise characteristics under unified weights greatly reduces the difficulty of system deployment in complex actual scenarios.

# Chapter 6
# Conclusions and Future Work

## 6.1 Conclusions

This paper designs to solve the limitations of the traditional active noise control (ANC) system in handling broadband unsteady noise and voice retention. We have proposed and verified a set of Deep ANC systems with selective noise cancellation capabilities. By building a simulation environment based on physical acoustics and designing a deep convolutional recurrent network (CRN), this study has made substantial breakthroughs in the following three aspects:

**First, it breaks through the physical limits of the linear control algorithm.**

The traditional FxLMS algorithm is limited by linear assumptions and is difficult to cope with the nonlinear distortion introduced by the speaker and the rapidly changing unstable noise. The Deep ANC system proposed in this study successfully realises the implicit compensation for hardware nonlinearity by using the powerful nonlinear mapping ability of the CRN architecture. The experimental results show that when dealing with the challenging non-steady Babble noise, the noise reduction of Deep ANC reaches 18.02 dB, which is more than 16 dB higher than FxLMS, which proves the dominant advantage of the deep learning model in complex acoustic environments.

**Second, the spectral-based selective processing of noise and speech is realised.**

In view of the traditional ANC indiscriminate elimination of sound limitations, this study introduces the speech retention loss function. This mechanism gives the system the ability to distinguish signals based on spectrum characteristics to ensure that the voice is completely preserved while suppressing noise. The test shows that the PESQ and STOI scores of the target voice after turning on noise reduction have been improved, which successfully solves the conflict between noise reduction and voice fidelity.

**Third, it verifies the generalisation robustness of the model in a reverberation environment.**

Unlike the existing research, which is mostly based on the ideal anechoic chamber hypothesis, this study uses the Image Source Method (ISM) to build a sound field with real reverb ($RT_{60} = 0.3s$). Experiments confirm that the Deep ANC system can effectively suppress multi-path reverberation interference thanks to the modelling ability of LSTM units for long-term dependencies. In addition, in the face of noise types (such as Volvo and F16) with distinct spectral and temporal characteristics, the system still maintains a robust noise reduction performance and shows a strong generalisation ability.

To sum up, the Deep ANC system has laid a solid technical foundation for the next generation of intelligent listening and AR audio equipment with its high efficiency, robustness and speech preservation capability.

## 6.2 Recommendation in Future Work

Although this research has achieved encouraging results in a single-channel simulation environment, it still needs to be deeply explored in the following key directions to truly implement the Deep ANC system into complex real scenarios:

1. **Exploration of time domain architecture for low latency**

Strict delay constraints are the biggest challenges facing the actual deployment of ANC. The frequency domain STFT method used in this study inevitably introduces frame-level latency (millisecond level), which is difficult to meet some ultra-low latency scenarios with extremely high requirements for phase matching (usually < 100 $\mu s$). Future research should turn to time-domain deep learning architecture (such as Wave-U-Net). The path completely eliminates the delay caused by the frequency domain transformation by direct point-to-point processing of the waveform. In addition, it is also necessary to combine model quantisation and pruning technology to further compress the computing overhead to adapt to the computing power limitations of embedded chips.

2. **Multi-channel and spatial selective expansion**

This study is currently limited to single-channel control, and the noise source in reality is often multi-faceted. Future research should expand the system into a multi-channel architecture. Combined with the concept of spatial selectivity proposed by Xiao et al. and using the beamforming ability of the microphone array [32], Deep ANC is expected to achieve accurate capture or suppression of sound sources in a specific direction. This means that the system can not only distinguish sounds according to spectral characteristics, but also filter sounds according to the direction of the source, further enhancing its ability to separate in a complex environment.

**3. Hybrid control architecture**

Although the pure end-to-end deep learning model performs well, the traditional adaptive filter has an extremely high energy efficiency ratio and stability when dealing with simple steady-state noise. Future research can refer to the hybrid system ideas proposed by Park et al. to combine traditional algorithms with deep neural networks [23]. Use traditional algorithms to process low-frequency steady-state components, while the deep network focusses on processing complex unsteady and nonlinear components. This complementary architecture is expected to find a better balance between computing cost and noise reduction performance.

**4. Adaptive mechanism of dynamic acoustic environment**

The current model is trained under a fixed reverberation time. However, the user's acoustic environment (such as moving from the office to the outdoors) is dynamically changed, which will cause the secondary path $S(z)$ to change. Future work can explore meta-learning and online fine-tuning technology. By introducing these mechanisms, the model will be able to sense environmental changes in real time and dynamically adjust parameters, so as to maintain stable performance in unknown sound fields [26].

Using Attentive Recurrent Network," in IEEE/ACM Transactions on Audio, Speech, and Language Processing, vol. 31, pp. 1114-1123, 2023.

[23] J. Park, H. H. Cho, H. Lim, and S. W. Lee, "HAD-ANC: A hybrid system comprising an adaptive filter and deep neural networks for active noise control," in Proceedings of Interspeech, Dublin, Ireland, pp. 5321-5325, August, 2023.

[24] R. Liu, Q. Liu, Y. Wang, W. Yu, and G. Cheng, "Neural network-based ANC algorithms: a review," Journal of Vibroengineering, vol. 27, no. 6, pp. 1105-1123, Aug. 2025.

[25] M. Solanki and A. Dusane, "A comprehensive review on active noise reduction methods for aircraft aerodynamics system," Journal of Vibration Engineering & Technologies, vol. 14, Art. no. 33, January, 2026.

[26] Z. Jiang, H. Xue, H. Yue, X. Bao, J. Zhu, X. Wang, and L. Zhang, "A review of artificial intelligence-driven active vibration and noise control," Machines, vol. 13, Art. no. 946, October, 2025.

[27] O. Yaish, Y. Mishaly, and E. Nachmani, "Active speech enhancement: Active speech denoising declipping and dereverberation," arXiv preprint arXiv:2505.16911, 2025.

[28] Y. Mishaly, L. Wolf, and E. Nachmani, "Deep active speech cancellation with Mamba-masking network," arXiv preprint arXiv:2502.01185, 2025.

[29] J. B. Allen and D. A. Berkley, "Image method for efficiently simulating small-room acoustics," Journal of the Acoustical Society of America, vol. 65, no. 4, pp. 943–950, April, 1979.